\documentclass[12pt]{iopart}
\usepackage{iopams}
\expandafter\let\csname equation*\endcsname\relax
\expandafter\let\csname endequation*\endcsname\relax
\usepackage{amsmath}
\usepackage{physics}
\usepackage[alsoload=synchem]{siunitx}
\usepackage{graphicx}

\newcommand{\kboltz}{k_{\mathrm{B}}}
\newcommand{\Deriv}[2]{\frac{\rmd#1}{\rmd#2}}

\newcommand{\eps}{\varepsilon}

\begin{document}
\title[Electrostatic enhancement factor for the coagulation of silicon nanoparticles]{Electrostatic enhancement factor for the coagulation of silicon nanoparticles in low-temperature plasmas}
\author{Benjamin Santos$^1$, Laura Cacot$^{1, 2}$, Claude Boucher$^1$, François Vidal$^1$}
\address{$^1$INRS - Énergie Matériaux Télécomunications, Varennes, QC J3X 1P7, Canada}
\address{$^2$Universit\'e Toulouse III - Paul Sabatier, 31062 Toulouse, France}
\ead{benjamin.santos@emt.inrs.ca}

\begin{abstract}
    The coagulation enhancement factor due to electrostatic (Coulomb and polarization-induced) interaction 
    between silicon nanoparticles was numerically
    computed for different nanoparticle sizes and charges in typical low-temperature argon-silane plasma conditions. 
    We used
    a rigorous formulation, based on a multipole moment coefficients, to describe the complete electrostatic
    interaction 
	between dielectric particles. The resulting interaction potential is non-singular at
	the contact point, which allows to adapt the orbital-motion limited theory to calculate the
	enhancement factor. It is shown that, due to induced polarization,
	coagulation is enhanced in neutral-charged particles encounters up to several orders of magnitude. 
	Moreover, the short-range force between like-charged nanoparticles can become attractive as a direct consequence 
	of the dielectric nature of the nanoparticles. 
	The multipolar coefficient potential is compared to an approximate analytic form
	which can be readily used to simplify the calculations.
	The results presented here provide a
	better understanding of the electrostatic interaction in coagulation and can be used in dust growth simulations 
	in low-temperature plasmas where coagulation is a significant process.

\end{abstract}

\maketitle
%
%
\section{Introduction}
  Nanodusty plasmas are composed of electrons, neutral and ionized particles of a gas and nanometric sized grains
  of condensed matter, that we will call nanoparticles. The growth of silicon nanoparticles in 
  low-temperature radio frequency capacitively coupled plasmas (RF-CCP) in an argon-silane (Ar-SiH$_4$)
   mixture has been the subject of several experimental
  investigations~\cite{boufendi_particle_1994, hollenstein_physics_2000,boufendi_dusty_2011}.
  In this paper, we limit the study to this configuration.
  Dusty plasmas are, however, ubiquitous in nature, in particular in space, and can also be generated over different
  laboratory conditions of gas mixture, temperature, pressure, and type of discharge~\cite{fortov_complex_2005}.

  The evolution of nanoparticle size starts with the nucleation
  phase~\cite{boufendi_particle_1994,hollenstein_physics_2000,boufendi_dusty_2011}, where primary dust is
  formed from the polymeric assembly of small molecules. Then follows the coagulation or agglomeration phase, which
  starts
  when a critical nanoparticle density is
  reached. In that phase, nanoparticles coalesce to form bigger particles, which can grow to tens of nanometers. This
  phase is identified by an increase in the size of the particles and a
  decrease in the nanoparticle number density.
  Then the surface growth, in which small molecules stick on the surface of the nanoparticles, becomes the more
  important mechanism of growth.

  It is known that nanoparticles in low-temperature Ar-SiH$_4$ plasmas are mostly charged negatively due
  to the high-frequency electron bombardment.
  Recently, Mamunuru et al.~\cite{mamunuru_existence_2017} have pointed out the existence of positively
  charged and neutral nanoparticles, along with the expected negatively charged particles. 
  Such a charge distribution
  promotes the coagulation because of the Coulomb attraction
  between particles of opposite charge polarity and the polarization-induced attraction taking place between
  all particles, but mostly between a neutral and a charged dielectric particle.

  This polarization-induced attraction force results from the polarization of the bound 
  charges residing in one particle induced by the electric field due to the presence of a net charge in the second
  particle. Several mathematical approaches have been used to calculate this force between two dielectric spherical
  particles (see for instance \cite{lindgren_progress_2016} and references therein).
  The ratio between the coagulation rate due to the electrostatic forces and the coagulation rate 
  of two neutral particles is called the electrostatic enhancement factor. 

  Ravi et al.~\cite{ravi_coagulation_2009} have studied theoretically the
  coagulation enhancement between neutral and charged silicon nanoparticles by using the approximate image 
  potential proposed by Huang et al.~\cite{Huang1991191} and the Amadon and Marlow's 
  expression~\cite{amadon_cluster-collision_1991} to calculate the enhancement factor. Both, van der Waals 
  and electrostatic charge-like interaction
  have not been taken into consideration in their treatment. They concluded that the coagulation enhancement due to
  induced polarization can be very significant.
  For instance, they found that the neutral-charged enhancement factor can be higher than the oppositely charged under
  certain conditions (see figure 1 in~\cite{ravi_coagulation_2009}).
  Using this model, they performed extensive self-consistent nanodusty plasma simulations to investigate nanoparticle
  growth.

  The use of point charges in the approach of Huang et al. to approximate the potential is, however, clearly inadequate
  when the sizes of the nanoparticles are comparable. Moreover, the divergence of this potential at the
  contact point between coalescing particles does not seem to have any physical ground.
  In addition, the complex Amadon and Marlow's expression for the enhancement factor, which was designed 
  to handle such singular potentials, was found to provide higher coagulation enhancements as
  compared to some alternative forms~\cite{ouyang_nanoparticle_2012}.

  These considerations motivated a revision of the calculation of the enhancement factor due to the electrostatic
  interaction between dielectric particles. In this work, we compute this enhancement factor 
  between two dielectric spheres by using the multipolar coefficient potential (MCP) of Bichoutskaia et 
  al.~\cite{bichoutskaia_elena_electrostatic_2010},
  which we consider more rigorous and appropriate to tackle this problem than the potential of Huang et al.
  Since the MCP of Bichoutskaia et al. is not singular at the contact point, a simple adapted
  orbital motion limited (OML) theory~\cite{allen_probe_1992} can be used to calculate the enhancement factor.

  Besides the electrostatic interaction, the van der Waals interaction is also known to play a significant
  role in the coalescence of particles \cite{Hamaker_1937}. The consistent treatment of the latter
  in our particular context is, however, outside the scope of this paper, which focuses essentially on the
  electrostatic interaction (see \cite{Priye_2013} for an overview of the general problem of the van der Waals
  interaction).

  This article is organized as follows. First, in \sref{sec:met1} we present the parameters and assumptions 
  of the following calculations.  Next, in \sref{sec:crate}
  we express the coagulation rate along with the enhancement factor. Their derivations are developed in the
  appendix. Then, in \sref{sec:limits}, we show the most probable equilibrium charge for a given nanoparticle size
  under the conditions of interest. In section \ref{sec:pot} we describe the MCP used to compute the interaction
  between the
  particles. Computational and numerical details are given in~\sref{sec:num}.
  Numerical results and their discussion are reported in~\sref{sec:res}. Finally, we present the conclusions
  in~\sref{sec:con}. 
\section{Methods}\label{sec:met}
\subsection{Plasma parameters and simplifying assumptions} \label{sec:met1}
  Typical parameters in low-temperature low-pressure RF-CCP are shown in \tref{tab:experiments}.
  \begin{table}
  \caption{Nanodusty plasma parameters}
  \begin{indented}
  \item[]\begin{tabular}{@{}ll}
  \br
  Parameter& value\\
  \mr
  Pressure $(p)$& \SI{100}{\milli \torr}\\
  Mean electron energy $\left(\varepsilon\right)$ & \SI{3}{\electronvolt}\\
  Gas/ion/nanoparticle temperature $\left(T\right)$ & \SI{300}{\kelvin}\\
  Ion density $\left(n_+\right)$& \SI{e15}{\meter^{-3}}\\
  Electron density $\left(n_e\right)$& \SI{9e14}{\meter^{-3}}\\
  Nanoparticle density $\qty(N_\text{p})$ & \SI{e14}{\meter^{-3}}\\
  Nanoparticle diameter $\qty(d_\text{p})$ & \SIrange[range-units=single]{1}{100}{\nano\metre}\\
  Nanoparticle mean free path $\qty(\lambda =1/\sqrt {2}N_p \pi d^{2}_p)$ & \SI{> 200}{\milli\meter}\\
  Dusty Debye length $\qty(\lambda_\text{D})$ & \SI{3.77e-5}{\meter}\\
  Nanoparticle charge $\qty(q)$ & \SIrange[range-units=single]{1}{296}{e}\\
  Interparticle distance $\qty(d=\qty(3/4\pi N_p)^{1/3})$ &  \SI{1.34e-5}{\meter}\\
  Structure parameter $\qty(\kappa_\text{d} = d/\lambda_{\text{D}})$ &  \SI{0.36}{}\\
  Knudsen number $\qty(\K_\text{n} = \lambda/d_p)$ & $\gg 1$\\
  Coupling parameter $\qty(\Gamma_\text{d})$ & \SIrange[range-units=single]{0.003}{254.316}{}\\
  \br
  \end{tabular}
  \end{indented}
  \label{tab:experiments}
  \end{table} 
  From these parameters, it is possible to deduce the following conditions:
  \begin{itemize}
  \item \textbf{Free molecular regime prevails.} At low pressure and low gas density, neutral nanoparticles move
  ballistically and behave like hard spheres $(\lambda \gg d_\text{p}, K_\text{n}\gg 1)$. As a consequence, only
  binary collisions are relevant.
  \item \textbf{Screening by ions is negligible.} The dusty Debye length
  is on the order of the ion Debye length
  $\lambda_\text{D}\sim\lambda_i$~\cite{shukla_colloquium_2009}.
  For the selected parameters, $\lambda_\text{D}$ is much greater than the nanoparticle sizes and greater than the 
  interparticle distance $d$. This allows using the basic form or the OML theory for nanoparticle charging and
  interaction.
  \item \textbf{Nanoparticle collective effects are negligible.}
  These effects are characterized by the coupling parameter 
  $\Gamma_\text{d}= \frac{1}{4\pi \eps_0}\frac{q^2}{\kboltz T d}\exp(-\kappa_\text{d})$, which is defined as the ratio
  between the Coulomb interaction energy and the nanoparticle thermal 
  energy~\cite{shukla_colloquium_2009}. This coupling
  is relatively low for this system. Crystallization is however expected for
  $\Gamma_\text{d}>170$~\cite{shukla_colloquium_2009}, which would occur only
  for particles having high charge states ($q > 290e$), corresponding to the largest nanoparticles considered in this
  work, i.e., $d_p \approx 100$ nm.
  As a consequence, we assume that the nanoparticles are not strongly interacting to form a crystal and that
  classical electrostatics can be used to describe the interaction between two particles inside the Debye
  sphere~\cite{fortov_dusty_2004}.
 \end{itemize}

  In addition, we make the following simplifying assumptions:
  \begin{itemize}
  \item \textbf{The average nanoparticle charge is provided by the argon ion and electron currents.} 
  While many ion species can coexist in Ar-SiH$_4$ plasmas, in the absence of a plasma chemistry model, we assume that
  the average nanoparticle charge is mainly given by the argon ion and electron currents, as given by the OML theory
  and electron tunneling.
  \item \textbf{Nanoparticles are spherical and have bulk silicon properties}. 
  As required by the MCP potential~\cite{bichoutskaia_elena_electrostatic_2010}, the charging models and
  the coagulation rate model used in this work, nanoparticles are treated
  as spheres~\cite{boufendi_particle_1994}. We also assume that they have a relative permittivity
  $\varepsilon = 11.68$, electron affinity $A_\infty = \SI{4.05}{\electronvolt}$~\cite{picard_effect_2016}, and a mass
  density $\rho_p = \SI{2330}{\kilo\gram .
  \metre^{-3}}$\cite{haynes2016crc}
  as in bulk silicon at \SI{300}{\kelvin}. These are the only parameters considered related to 
  the nanoparticle composition.
  \item \textbf{The charge is uniformly distributed at the surface of the nanoparticles}. 
   As in our system, dielectric nanoparticles are being bombarded by the electrons and ions of the plasma at a rapid
   rate from all directions, we assume, consistently with the tunnel effect and OML theory for charging, that the
   charges are uniformly distributed at the surface of the nanoparticles.
  \end{itemize}
\subsection{Coagulation rate and enhancement factor}\label{sec:crate}
  The coagulation rate (\SI{}{\meter^{-3} . s^{-1}}) for two colliding spherical particles $i$ and $j$, with
  densities $n_i$ and $n_j$ (\SI{}{\meter^{-3}}), is expressed as:
  \begin{equation}
    R = \beta_{ij} n_i n_j,
    \label{eq:coagrate}
  \end{equation}
  where $\beta_{ij}$ (\SI{}{\meter^{3} . s^{-1}}) is the total coagulation kernel. Thus, when two particles
  of volume $v_i$ and $v_j$ coalesce,
  they form a new particle with volume $v_k$ which is equal to the sum of the two constituent particles,
  i.e. $v_k = v_i+v_j$.
  In the case of discrete particle sizes, the dynamics of coagulation is governed by
  the Smoluchowski equation~\cite{friedlander_smoke_2000}:
  \begin{equation}
    \Deriv{n_k}{t}  = \frac{1}{2}\sum_{i+j = k}\beta_{ij} n_i n_j - n_k\sum^\infty_{i=1}\beta_{ik}n_i,
    \label{eq:smol}
  \end{equation}
  where the first term of the right side represents the creation rate of the particle $k$ from
  the collision between the particles $i$ and $j$, and the second term
  is the annihilation of $k$ when it collides with any particle $i$.

\begin{figure}\centering
  \includegraphics[scale=0.6]{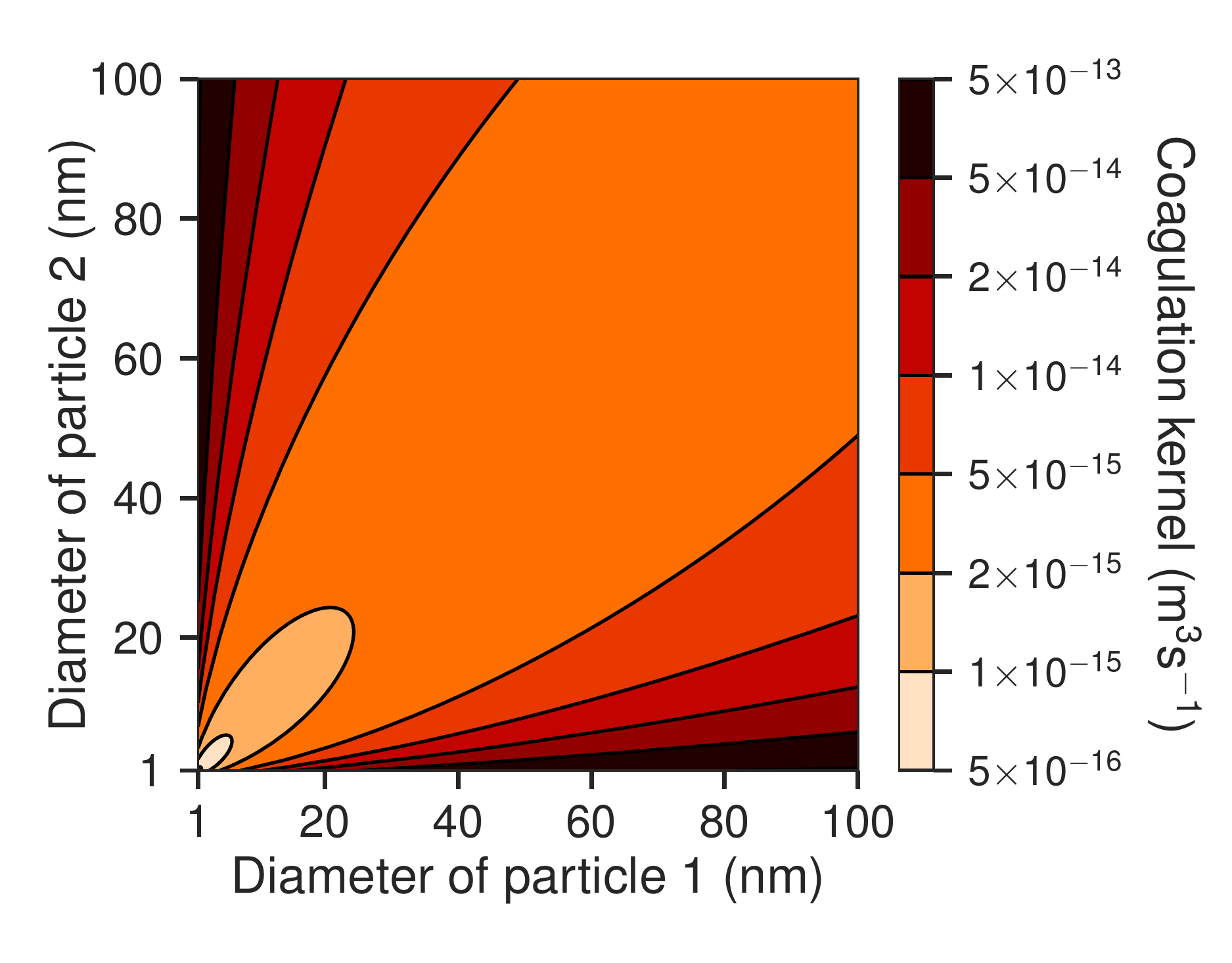}
  \caption{Coagulation kernel~\eref{eq:neutralkernel} in the free molecular regime as a
  function of
  coalescing particle sizes for $T=$\SI{300}{\kelvin} and $\rho_p$=\SI{2330}{\kg/\metre^3}.}
  \label{fig:beta_free}
  \end{figure}

  The total coagulation kernel can be separated into two factors,
  \begin{equation}
    \beta_{ij} = \eta_{ij}\beta^{0}_{ij}.
    \label{eq:totalkernel}
  \end{equation}
 where $\eta_{ij}$ is the enhancement factor, which accounts for the electrostatic interaction between
  nanoparticles. $\beta^0_{ij}$ is
  the coagulation kernel (or rate coefficient) for neutral particles in the free molecular regime, 
  which can be expressed as~\cite{friedlander_smoke_2000, ravi_coagulation_2009}:
  \begin{equation}
    \beta^{0}_{ij} = \left(\frac{3}{4\pi}\right)^{1/6}
                     \left[\frac{6 \kboltz T}{\rho_p}\left(\frac{1}{v_i}+\frac{1}{v_j} \right)\right]^{1/2}
                     \left(v^{1/3}_i + v^{1/3}_j \right)^2,
    \label{eq:neutralkernel}
  \end{equation}
  where $\kboltz$ is the Boltzmann constant, $T$ is the
  kinetic temperature of the nanoparticles, and  $\rho_{p}$ is their mass density. \Fref{fig:beta_free} shows
  the coagulation rate for spherical particles in the range of sizes of interest. One observes that
 coagulation between small and big particles is more efficient than between particles of the same size.
  This situation typically occurs in nanodusty plasmas where there are small nucleated
  particles with $d\sim$~\SI{1}{\nano \metre} and particles grown by coagulation and 
  surface growth with $d\sim$~\SIrange[range-units=single]{20}{100}{\nano \metre}~\cite{ravi_coagulation_2009}.

  Within the framework of the OML theory~\cite{allen_probe_1992}, the enhancement factor
  can be written explicitly (see~\ref{app:derivrate}),
  \begin{equation}
   \eta_{ij} = \exp\left(-\frac{\Phi_{ij,\text{max}}}{\kboltz T}\right)
      \left[ 1 + \frac{\Phi_{ij,\text{max}}-\Phi_{ij} (r_\text{min})}{\kboltz T} \right].
    \label{eq:etagen}
  \end{equation}

  In this expression, $r_\text{min} = r_i + r_j$ is the distance between the centers
  of the spherical particles $i$ and $j$, of radius $r_i$ and $r_j$, respectively, when they are in contact,
  and $\Phi_{ij,\text{max}}$ is the maximum value of the interparticle potential (see~\sref{sec:pot}).
  Obviously, $\eta_{ij} = \eta_{ji}$ since $\Phi_{ij}=\Phi_{ji}$.
  
  Two important particular cases can be distinguished:
  \begin{itemize}
    \item monotonically increasing (attractive) potential ($\Phi_{ij}(r_{\text{min}}) < 0, \Phi_{ij,\text{max}} = 0$):
      \begin{equation}
        \eta_{ij} = 1 - \frac{\Phi_{ij} (r_\text{min})}{\kboltz T}, \qquad
          \Phi_{ij}(r) \leq 0\, \forall\, r \geq r_\text{min},
        \label{eq:etaattractive}
      \end{equation}
    \item monotonically decreasing (repulsive) potential ($\Phi_{ij,\text{max}} = \Phi_{ij}(r_\text{min})>0$):
      \begin{equation}
        \eta_{ij} = \exp\left(-\frac{\Phi_{ij} (r_\text{min})}{\kboltz T}\right), \qquad
          \Phi_{ij}(r) \geq 0\, \forall\, r \geq r_\text{min}.
        \label{eq:etarepulsive}
      \end{equation}
  \end{itemize}
  Naturally, these last two cases arise for instance when $\Phi_{ij}$ is the Coulomb potential for point 
  charges~\cite{shukla_colloquium_2009, allen_probe_1992}. In the present situation, where induced polarization
  exerts an attractive force between particles, case \eref{eq:etaattractive} 
  is met for neutral-charge and oppositely charged particles, while either case \eref{eq:etarepulsive}
  or $\Phi_{ij, \text{max}} > 0$ and $\Phi_{ij} (r_{\text{min}}) < \Phi_{ij, \text{max}}$ can be met for like-charged
  particles.
\subsection{Equilibrium charge and charge distribution width}\label{sec:limits}
  Two regimes of nanoparticle charging can be identified. In the first, the negative charge is limited by the
  tunnel effect, making the occurrence of small positively charged particles possible~\cite{heijmans_comment_2016,
  mamunuru_existence_2017}. In the second regime, the mean charge and the charge distribution
  width are inferred from the balance between the electronic and ionic currents towards the particle,
  as given by the OML theory~\cite{allen_probe_1992}. The OML collision frequencies for the plasma species 
  $j=e$, Ar$^+$ charging the nanoparticle of radius $r_i$ and charge $q_i$, is~\cite{allen_probe_1992}:
    \begin{equation*}
      \nu_{ji}= 4\pi r^2_i n_j \left( \frac{E_j}{2\pi m_j}\right)^{1/2}\alpha_{ji},
    \end{equation*}
  where $n_j$ and $m_j$ are the number density and mass, respectively. $E_j$ is the average kinetic energy, 
  specifically $E_e = \varepsilon$ and $E_{\text{Ar}^+}= 3k_BT/2$, according to \tref{tab:experiments}. The coefficient
  $\alpha_{ji}$ and the surface nanoparticle potential $\phi_i$ are given by:
     \begin{align*}
      \alpha_{ji} &= \left \{
      \begin{array}{ll}
       \exp\left(-\frac{q_j\phi_{i}}{k_B T_j} \right) & \quad\textrm{for } q_j q_i \geq 0,  \\
       1-\frac{q_j\phi_{i}}{k_B T_j}  & \quad\textrm{for } q_j q_i < 0,
      \end{array} \right .\\
      \phi_{i}&\equiv \phi(r_i,q_i)=\frac{q_i}{4\pi\varepsilon_{0}r_i}.
     \end{align*}

  The charging current associated with the species $j$ is simply the product of the OML frequency
  and the charge, i.e. $I_{j,\text{OML}} = q_j\nu_{ji}$.

  The electron tunneling current is expressed as~\cite{heijmans_comment_2016, mamunuru_existence_2017, griffiths_wkb}:
  \begin{equation}
   I_{e, \text{tunnel}} = \frac{2\abs{q_i}}{\hbar} \left[\kboltz T e\phi(r_a, q_i)\right]^{\frac{1}{2}}
 \left[ \beta \cos^{-1} \left( \beta^{-\frac{1}{2}} \right) - (\beta-1)^{\frac{1}{2}} \right],
   \label{eq:itunnel}
  \end{equation}
  where it is assumed that $q_i \le 0$. In this expression, $\hbar$ is the reduced Planck constant,
  \begin{equation}
   \beta = \frac{r_a}{r_i} = \frac{q_i}{q_i + 5e/8 - 4\pi \varepsilon_0 r_i A_\infty/e},
   \label{eq.ra}
  \end{equation}
  where $A_\infty = \SI{4.05}{\electronvolt}$ is the electron affinity of the uncharged flat bulk silicon, and
  $r_a \ge r_i$ is the location
  where electrons must tunnel to leave the nanoparticle~\cite{mamunuru_existence_2017}. 
  This result uses the fact that, on a negatively charged nanoparticle, electrons lie in a potential well,
  characterized by the electron affinity $A_\infty$, while outside the well electrons undergo the repulsive Coulomb
  force. The current \eref{eq:itunnel} expresses the probability for these electrons to pass
  through the barrier.

\begin{figure}\centering
    \includegraphics[scale=0.6]{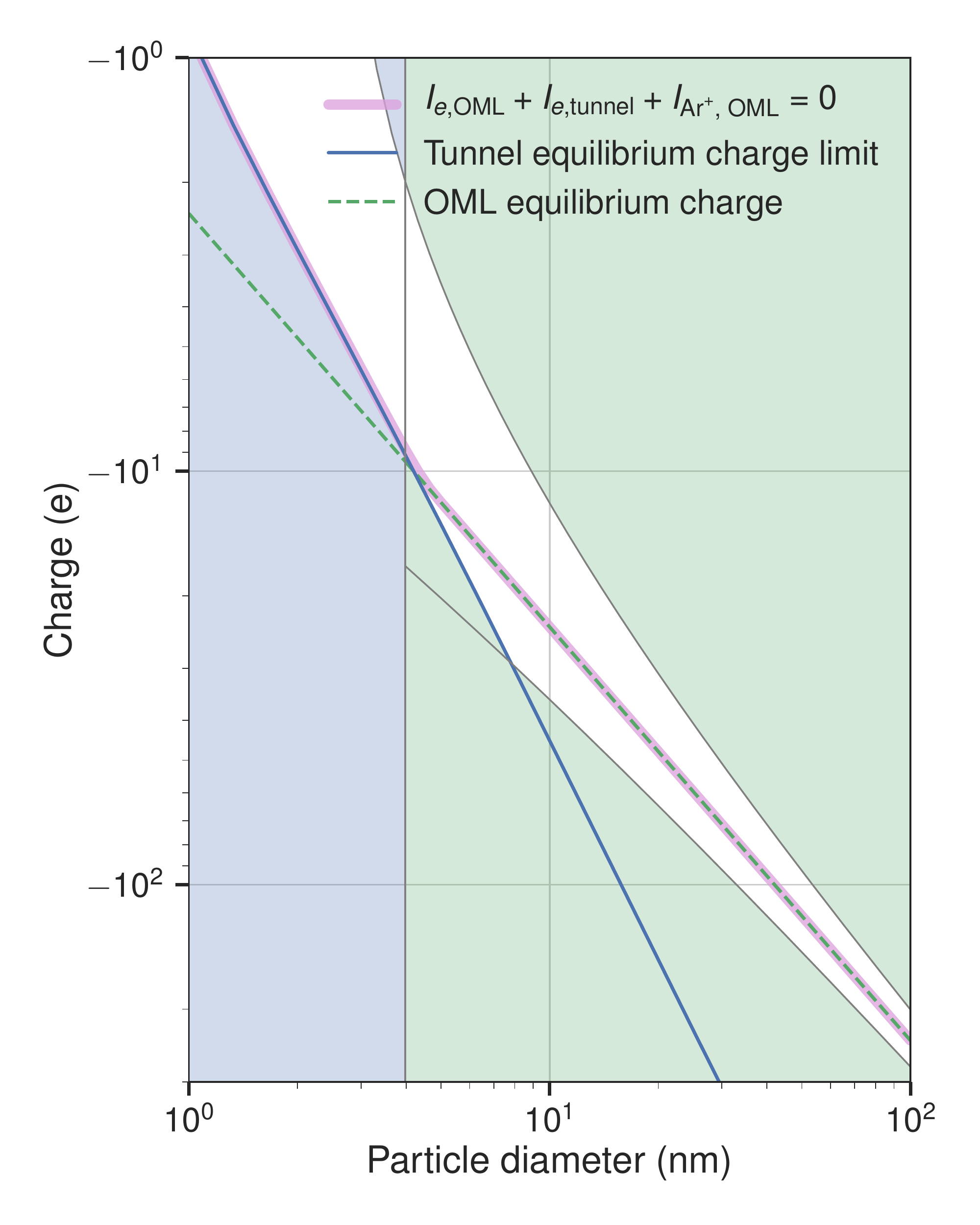}
    \caption{Equilibrium charge and charge distribution width as a function of particle diameter. The
    equilibrium charge in the tunnel regime is represented by the blue solid line and is considered as the
    maximum negative charge that a particle can bear when it is greater than the charge in the OML regime. 
    The equilibrium charge in the OML regime is given by the
    green dashed line enclosed by two lines which delineate the width of the charge distribution. The blue and
    green zones outside the white area denote less likely combinations of size and charge for nanoparticles
    close to equilibrium with the plasma. The equilibrium charge resulting from the three currents is 
    represented by the pink curve.}
  \label{fig:charges}
  \end{figure}

  By balancing the currents, $I_{e,\text{OML}} + I_{e, \text{tunnel}} + I_{\text{Ar}^+, \text{OML}}=0$, 
  the equilibrium charge can be estimated as a function of the particle size. 
  Comparing the positive currents, it is possible to distinguish two regimes: 
  the tunnel regime, where $I_{e, \text{tunnel}} \gg I_{\text{Ar}^+,\text{OML}}$, for $d < 4$ nm,
  and the OML regime, where $I_{e, \text{tunnel}} \ll I_{\text{Ar}^+,\text{OML}}$, for $d > 4$ nm.
  This is shown in~\fref{fig:charges} for the plasma conditions given in~\tref{tab:experiments}. Since the tunnel
  current acts very fast, the equilibrium tunnel charge can be approximated as the maximum negative charge the particle
  can bear~\cite{picard_effect_2016, mamunuru_existence_2017, heijmans_comment_2016}.
  For a mean charge number $z$, the width of the charge distribution can be expressed as
  $z\pm 3\sigma$, where $\sigma^2$ is the variance of the charge distribution as given
  by the OML theory~\cite{picard_effect_2016}:
  \begin{equation}
   \sigma^2 = \frac{4\pi\varepsilon_0 r_i \kboltz T_e}{e^2}\left( 1-\frac{T_e}{T_e+T - 
   \frac{e^2}{4\pi\varepsilon_0 r\kboltz}z} \right).
   \label{eq:omlsigma}
  \end{equation}
\subsection{Interparticle potential}\label{sec:pot}
  Considering the uniform surface charge distribution implied by the tunnel effect and the OML theory for charging, the
  multipolar coefficients potential (MCP), developed by
  Bichoutskaia et al.~\cite{bichoutskaia_elena_electrostatic_2010, stace_reply_2012}, appears to be well suited for
  the problem of interest since their MCP deals with the electrostatic interaction of two dielectric particles 
  with charges distributed on their surface.
  The MCP was found to provide a well-behaved solution that agrees with the more general but more complicated solution
  of Khachatourian et al.~\cite{khachatourian_electrostatic_2014} expressed in terms of bispherical 
  coordinates. Furthermore, the MCP solution converges
  rapidly using fewer terms in the expansion than the bispherical solution~\cite{lindgren_progress_2016}.

  The MCP reads~\cite{bichoutskaia_elena_electrostatic_2010, stace_reply_2012},
  \begin{align}
    \Phi_{\text{MCP}}(r, r_i, q_i, r_j, q_j) & \equiv \Phi_{\text{MCP,ij}}(r)\\ & = K\frac{q_iq_j}{r}\nonumber\\
      &-\frac{q_i}{2}\sum^{\infty}_{m=1} \sum^{\infty}_{l=0}A_{l}
      \frac{\left(\varepsilon-1\right)m}{\left(\varepsilon+1\right)m+1}
      \frac{\left(l+m\right)!}{l!m!}\frac{r^{2m+1}_j}{r^{2m+l+2}}\nonumber\\
      &-\frac{1}{2K}  \sum^{\infty}_{l=1} A^2_{l}
      \frac{\left(\varepsilon+1\right)l+1}{\left(\varepsilon-1\right)lr^{2l+1}_i},
    \label{eq:pmc}
  \end{align}
  where $r$ is the distance between the centers of the particles,
   $q_i$ and $q_j$ are the charges of particles $i$ and $j$, respectively,
   and $K=1/4\pi\varepsilon_0$.
  The multipolar coefficients $A_{l}$, which takes into account the mutual polarization of the particles, are the
  solutions of the following linear system of equations,
  \begin{align}
    A_{j_1} &= Kq_1\delta_{j_1, 0}
    - \frac{\left( \varepsilon-1 \right)j_1}{\left( \varepsilon+1 \right)j_1+1}
      \frac{r^{2j_1+1}_1}{r^{j_1+1}}Kq_j\nonumber
      \\&+ \frac{\left( \varepsilon-1 \right)j_1}{\left( \varepsilon+1 \right)j_1+1}
      \sum^{\infty}_{j_2=0} \sum^{\infty}_{j_3=0}
      \frac{\left( \varepsilon-1 \right)j_2}{\left( \varepsilon+1\right)j_2+1}
      \frac{\left(j_1+j_2\right)!}{j_1!j_2!}\frac{\left(j_2+j_3\right)!}{j_2!j_3!}
      \frac{r^{2j_1+1}_i r^{2j_2+1}_j}{r^{j_1+2j_2+j_3+2}}A_{j_3}.
      \label{eq:mc}
  \end{align}
  
  In practice, the sums in equations~\eref{eq:pmc} and~\eref{eq:mc} are truncated with maximum indices determined
  as explained below. 
\subsection{Outline of the numerical calculations}\label{sec:num}
  In order to find the enhancement factor \eref{eq:etagen} using the MCP, 
  it is necessary to solve the linear system of equations
  \eref{eq:mc} and calculate the potential \eref{eq:pmc} at various relative distances $r$. This procedure can be quite
  time-consuming depending on the size of the linear system of multipolar coefficients.
  We determined the number of terms retained in the summations in equations~\eref{eq:pmc} and~\eref{eq:mc}
  from the convergence rate of the series as a function of the number of terms considered, knowing that the
  series converges uniformly. 
  If few consecutive values of the potentials with decreasing number of terms were in agreement within 1\% or less, 
  we considered that the number 
  of terms was sufficient. Otherwise, we raised the number of terms until the required convergence rate is achieved.
  The number of terms satisfying this condition is between 25 and 500, depending on the MCP arguments.

  Calculations were done using C++ codes complemented by the Boost library for solving linear systems of
  equations and finding roots~\cite{boostmath}.
  Additional calculations were performed in 
  Python with the scipy package~\cite{scipy} (results in~\ref{app:mpc-ipa-comparison} 
  and~\ref{app:maximum-image-potential-approximation-ipa}).
  The results were plotted using matplotlib~\cite{Hunter_2007}.
  More details can be found in the supplementary material from 
  FigShare~(https://doi.org/10.6084/m9.figshare.c.4206779).

  The calculation procedure can be summarized as follows:
  \begin{enumerate}
   \item Set the nanoparticle kinetic temperature, relative permittivity, sizes and charges pairs $(r_i, q_i)$.
   \item For each combination of pairs $(r_i, q_i)$ and $(r_j, q_j)$, compute the potential~\eref{eq:pmc} at
   the contact point
   $r_\text{min} = r_i+r_j$.
   \item In the case of like-charged particles, find the zero of the electrostatic force $-d\Phi_{ij}/dr$
   for $r > r_\text{min}$. If a zero force exists at a finite value of $r$, it corresponds to the potential barrier.
   Its height $\Phi_{ij,\text{max}}$ can thus be computed, and then the enhancement factor~\eqref{eq:etagen}.
   If such a root cannot be found, then the interaction is either purely repulsive or purely attractive
   and only $\Phi_{ij}(r_\text{min})$ is needed (cases \eqref{eq:etarepulsive} and \eqref{eq:etaattractive},
   respectively).
  \end{enumerate}

  The MCP can be expressed in terms of diameter and charge ratios $d_i/d_j$ and $q_i/q_j$, but the enhancement
  factor cannot be factorized that way. Thus, it is necessary to calculate the MCP
  for all distinct combinations of ($d_i,q_j$) and ($d_i,q_j$) of interest. To accelerate the calculations, the MCP was
  calculated for specific ratios and
  then extracted for all combinations of ($d_i,q_j$) and ($d_i,q_j$) to obtain the enhancement factor.
  For this work, we have considered $L=100$ diameters
  and $Q=302$ charges for each particle, which gives a total of $LQ(LQ+1)/2 \approx \SI{4.56e8}{}$ distinct
  combinations of pairs of coagulating particles. Only a subset of these combinations are, however, discussed
  in section~\ref{sec:res}.
  
  Furthermore, we have found that significant speedup can be achieved 
  to calculate $\Phi_{ij,\text{max}}$ by using an approximate formulation of the MCP, the Image Potential Approximation
  (IPA), which is discussed in~\ref{app:mpc-ipa-comparison}. The IPA provides an useful approximation of the MCP
  for all interparticle distances $r$ when the size of the particles are very different 
  and at large distance for any particle size ratios (see figure \ref{eq:IPA} for instance). 
  The IPA proves to be convenient to localize the position of $\Phi_{ij,\text{max}}$. With this hybrid approach, 
  the MCP needs to be computed no more than two times: at contact and at the barrier, for each combination of 
  $(r_i, q_i)$ and $(r_j, q_j)$ where $q_iq_j >0$. The full MCP was used, however, in this paper to calculate 
  $\Phi_{ij,\text{max}}$.
\section{Results and discussion}\label{sec:res}
\subsection{Coulomb enhancement factor}
\begin{figure}\centering
    \includegraphics[scale=0.45]{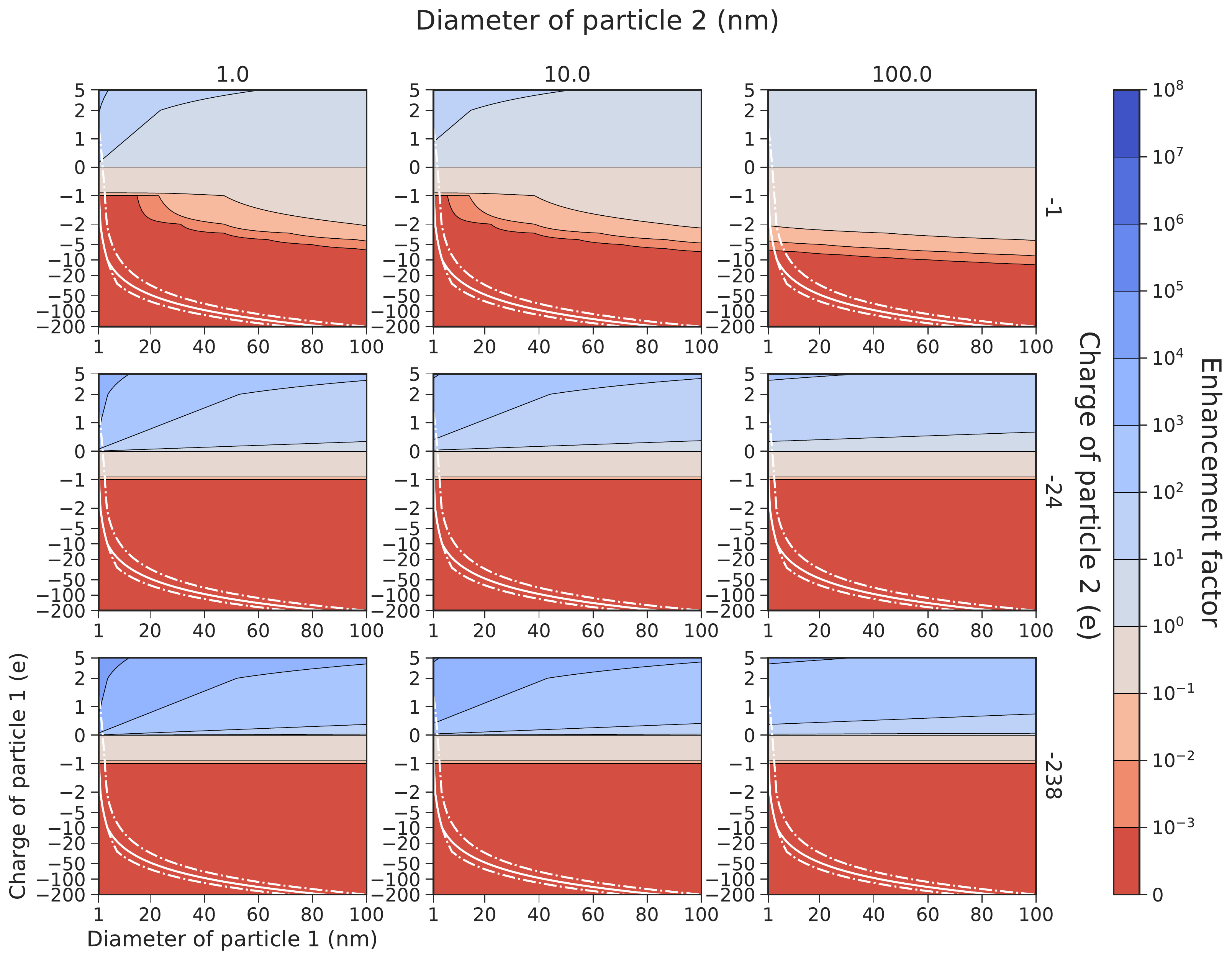}
    \caption{Contour levels of the Coulomb enhancement factor as a function of the
    size and charge of particle 1. Each row corresponds
    to a charge and each column to a diameter of particle 2. 
    The continuous white curve represents the most probable charge at equilibrium while the dot-dashed curves
    delimit the width of the charge distribution (see \fref{fig:charges}). Note that the particle charges are 
    discrete and the contours are being interpolated between the different discrete levels.}
  \label{fig:etacoul}
  \end{figure}

  For the sake of comparison with the MCP, we first present in \fref{fig:etacoul} the enhancement factor $\eta_{ij}$ 
  for the Coulomb interaction, as given by equations \eref{eq:etaattractive} and~\eref{eq:etarepulsive}.
  Each panel shows contour levels of the enhancement factor as a function of the size and charge of particle 1, whereas
  the size and charge of particle 2 are specified in columns and rows, respectively.
  The panels along the diagonal from upper left to lower right correspond to 
  equilibrium combinations of size and charge of particle 2 in the stated plasma conditions (see \fref{fig:charges}). 
  The size and charge of particles 1 and 2 outside the equilibrium values are also included to represent 
  possible, but less likely, states that could occur in dynamic out of equilibrium systems. 

  As expected, enhancement $\eta_{12}>1$ (blue zones) occurs for opposite charge particles
  and suppression $\eta_{12}<1$ (red zones) for like-charged particles. For the neutral-charged particles interaction,
  the enhancement factor is identically equal to 1 and is thus not represented.
  The enhancement factor shown in \fref{fig:etacoul} reflects the fact that the Coulomb interaction 
  decreases as $r_{\text{min}}$ increases for a given 
  charge $q_1$, and that attraction increases as $-q_1q_2$ increases.
  Since the white curve in \fref{fig:etacoul} represents the most likely charge at equilibrium (the same
  as in~\fref{fig:charges}),
  one can see that suppression occurs along this curve for all the selected pairs of size and charge of particle 2.

\begin{figure}\centering
    \includegraphics[scale=0.45]{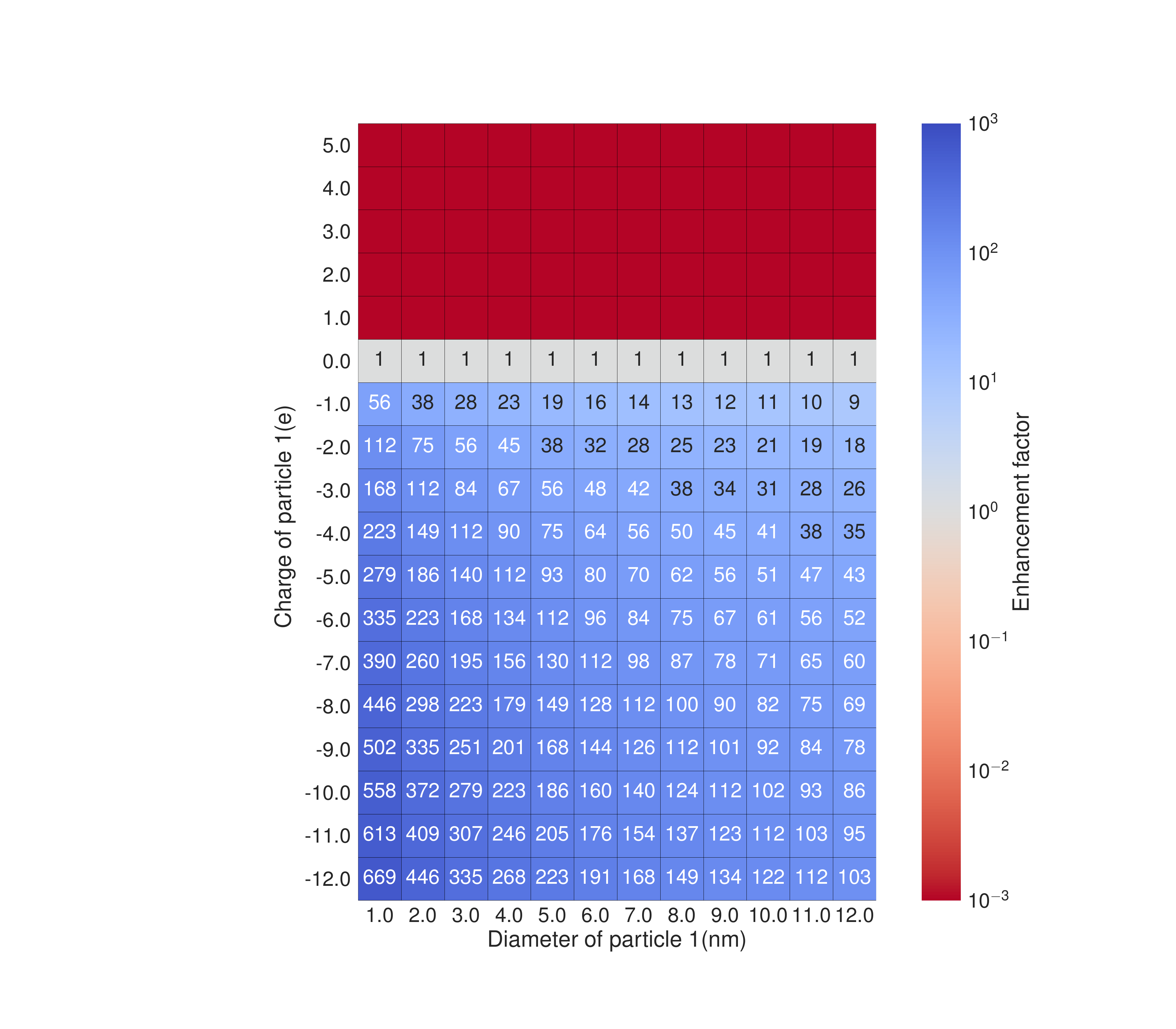}
    \caption{Chart of the Coulomb enhancement factor for small particles.
    The particle 2 has $d_1=1~\SI{}{\nano\metre}$ and $q_1=+e$. 
    Strong suppression occurs when charges have the same sign and the enhancement factor is exactly
    1 when the particle 1 is neutral.}
  \label{fig:etacoulsmall}
  \end{figure}

  For use in the following sections,~\fref{fig:etacoulsmall} shows a chart of the 
  enhancement factor as a function of the size and charge of small particles 1, for a particle 2
  of size $1~ \SI{}{\nano\metre}$ and charge $+e$.
\subsection{MCP charged-charged enhancement factor}\label{sec:charged}
  The MCP enhancement factor is shown in~\fref{fig:etapmc} for the same sizes and charges
  of particles 1 and 2 as those of \fref{fig:etacoul}. 
  Significant differences can be observed between the two cases. Red zones have recessed while blue zones
  have expanded in all panels, indicating a larger enhancement factor, which is a consequence of the
  polarization-induced attraction.
  
 \subsubsection{Opposite charge interaction.} 
 Contrary to the Coulomb interaction, the enhancement factor does not always decrease as  $d_1$ increases for
  opposite charge interaction, as can be seen in some panels for $q_2=-24e$ and $-238e$
  (horizontal sections for $q_1 > 0$). This effect is due to stronger
  induced polarization in larger particles. It is not observed for $q_2=-e$ because the charge is too small to induce 
  a significant polarization in particle 1. These trends can be understood by using the IPA, equation (\ref{eq:IPA}),
  which provides a useful approximation to the MCP at the contact point $r_{min}=(d_1+d_2)/2$ when the particle sizes
  $d_1$ and $d_2$ are very different, as this case is close to the one considered by Draine and
  Sutin~\cite{draine_collisional_1987} on which the IPA is based. The IPA at the contact point reads,
  to the first order in $d_2/d_1 \ll 1$,
  \begin{equation}
  \Phi_{\text{IPA},12}(r_{min}) \approx 2K \frac{q_1q_2}{d_1} \left(1-\frac{d_2}{d_1} \right) -\frac{K\kappa}{2}
  \frac{q_2^2}{d_2}\left(1-\frac{5}{2}\frac{d_2}{d_1}\right),
  \label{eq.PHIIPA}
  \end{equation}
 where the first term is the Coulomb interaction and the second term is the induced polarization. 
 From this expression, one finds that,
 for a given charge ratio $q_1/q_2 < 0$ and for size ratios such that $d_2/d_1 \ll 1$, the enhancement factor
 $\eta_{12}$, here given by equation (\ref{eq:etaattractive}), increases as $d_1$ increases if
  \begin{equation}
  -\frac{q_1}{q_2} < \frac{5}{8} \kappa.
  \end{equation}
 When this inequality is satisfied, the increase of the attractive force of the induced polarization is able
 to counterbalance the decrease of the Coulomb attraction as the size of the particle 1 increases.
 This condition works in all panels for $d_2=1$ nm and 10 nm, where $d_2/d_1 \ll 1$.

 For size ratios such
 that $d_1/d_2 \ll 1$, as is the case in the panels where $d_2=100$ nm, the condition for $\eta_{12}$ to increase
 as $d_1$ increases is,
 
 \begin{equation}
  -\frac{q_1}{q_2} <  \kappa \left[ \frac{3}{2}
  \left( \frac{d_1}{d_2} \right)^2
  - \left( \frac{q_1}{q_2}\frac{d_2}{d_1} \right)^2 \right].
  \end{equation}
  This condition can be satisfied only if the right-hand side is positive, namely
  \begin{equation}
    \left|\frac{q_1}{q_2} \right| <  \sqrt{\frac{3}{2}} \left(\frac{d_1}{d_2}\right)^2.
  \end{equation}
  Therefore the ratio $|q_1/q_2|$ must be very small and this is the case for $q_2=-238e$.
  
  In conclusion, in both limiting cases $d_2/d_1 \ll 1$ and
  $d_1/d_2 \ll 1$, the enhancement factor will increase as $d_1$ increases provided the charge ratio $-q_1/q_2$ is
  sufficiently small. Otherwise, the Coulomb attraction dominates the trend.
  
   \subsubsection{Like-charged interaction.} 
  For like-charge particles, one observes that the enhancement factor generally increases as $d_1$ increases, as in
  the Coulomb case, and this effect is enhanced due to the attraction induced by polarization.
  One exception is the $d_2=100$ nm, $q_2=-e$ panel where the opposite trend is observed. In this particular case,
  the induced polarization is large for small values of $d_1$ and decreases as $d_1$ increases for a sufficiently high
  charge ratio $q_1/q_2 >0$ (as also observed in~\cite{lindgren_progress_2016}).
  This can be readily seen from the IPA in the limit $d_1/d_2 \ll 1$,
  \begin{equation}
   \Phi_{\text{IPA},12}(r_{min}) \approx 2K \frac{q_1q_2}{d_2} \left(1-\frac{d_1}{d_2} \right)
   -\frac{K\kappa}{2} \frac{q_2^2}{d_2}
   \left[ \left( \frac{q_1}{q_2} \right)^2 \frac{d_2}{d_1}
   + 2\left( \frac{d_1}{d_2} \right )^3 \right],
  \end{equation}
  where the first term is the Coulomb interaction and the second term is the induced polarization.
  
  Moreover, contrary to the Coulomb interaction, the enhancement factor
  does not always increase monotonically as $-q_1q_2$ increases for a given $d_1$. For instance,
  in the $d_2=\SI{100}{\nano\metre}$,
  $q_2=-e$ panel, one can see that the enhancement factor at $d_1=\SI{40~}{\nano\metre}$ has a maximum near $q_1=-10e$,
  where $\eta_{12}>1$, although the repulsive Coulomb force is weaker near $q_1=-e$, where $\eta_{12}<1$. This effect
  is the result of peculiar combinations of $\Phi_{12,\text{max}}$, and $\Phi_{12}(r_{\text{min}})$ 
  entering in the
  expression of the enhancement factor, equation (\ref{eq:etagen}). In such cases, the IPA is of little use to
  understand the trends since the mathematical expressions are quite complicated.

\begin{figure}\centering
    \includegraphics[scale=0.45]{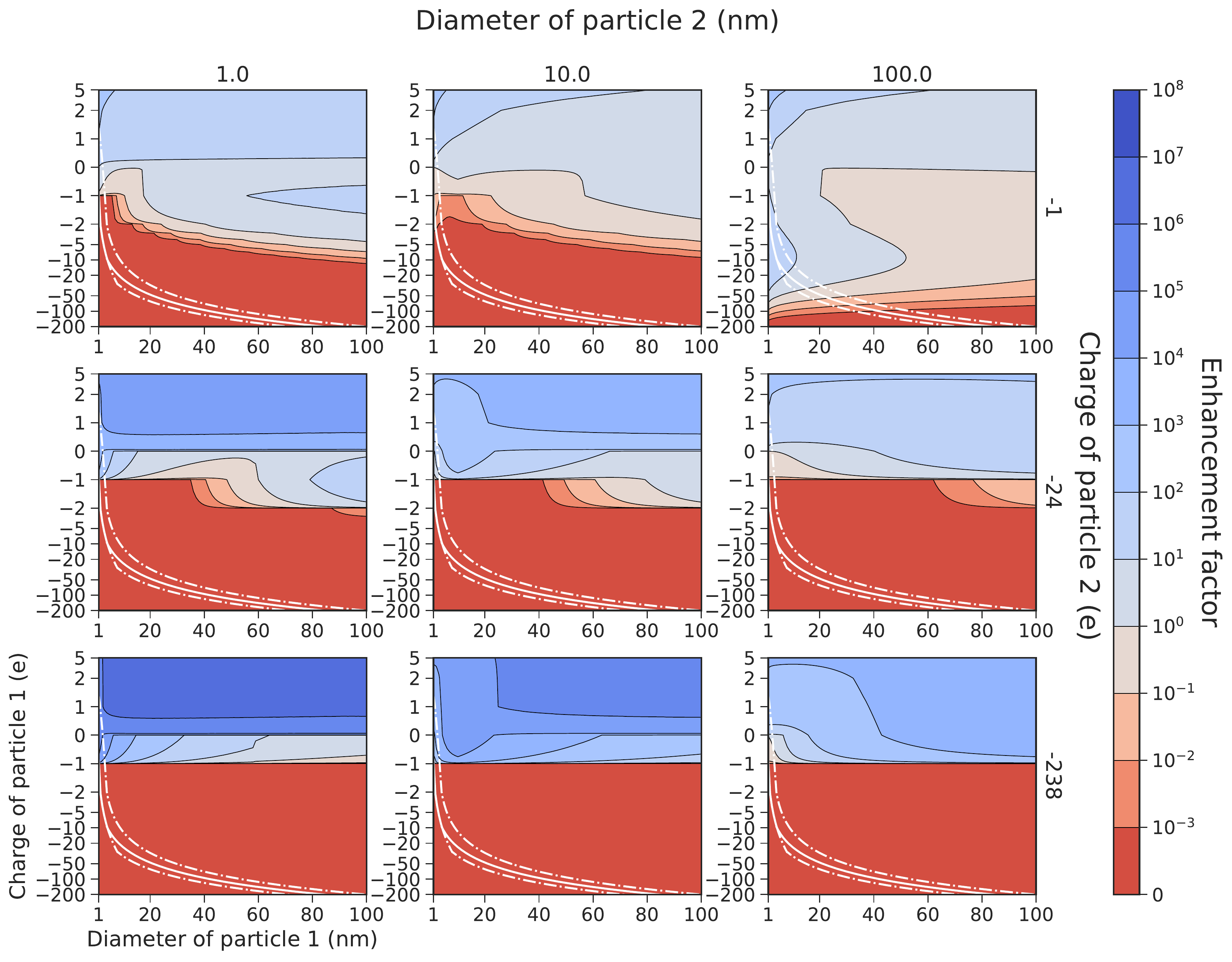}
    \caption{Contour levels of the MCP enhancement factor for particle sizes and charges as in~\fref{fig:etacoul}.}
  \label{fig:etapmc}
  \end{figure}

  The most striking features of \fref{fig:etapmc} is certainly that enhancement occurs for like-charged particles, as
  can be observed in panels for $q_2 =-e$, and $-24e$. For $q_2 =-238e$, the Coulomb repulsion is too strong for this
  effect to happen. The enhancement for like-charged particle interaction takes place, however, outside the equilibrium
  diameter-charge curve shown in~\fref{fig:charges}. For instance, large particles with a low charge 
  (e.g. $d > \SI{20}{\nano\metre}$ and $q \approx -e$), where enhancement is observed, are unlikely at equilibrium.

\begin{figure}\centering
    \includegraphics[scale=0.45]{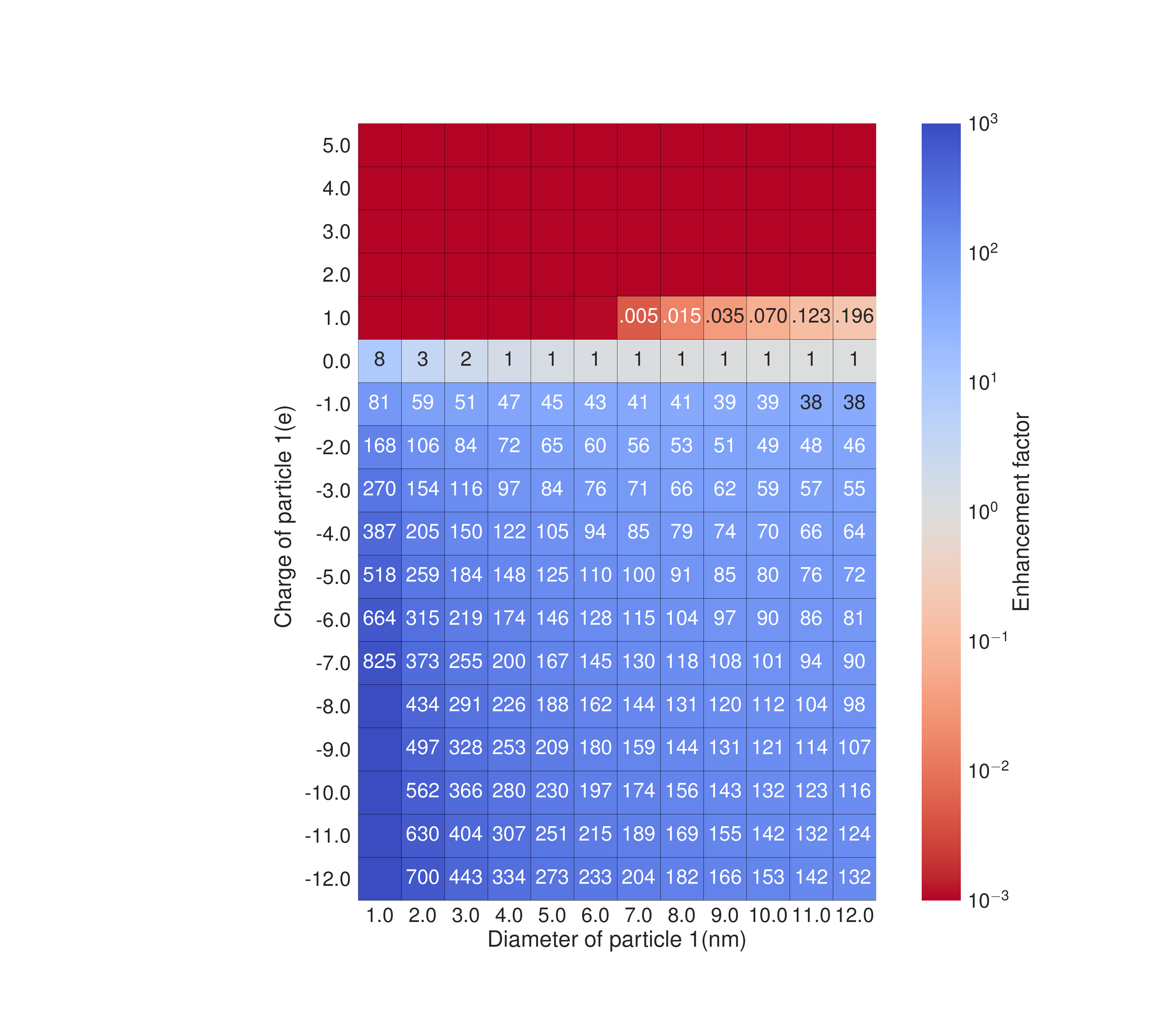}
    \caption{Chart of the MCP enhancement factor for small particles.
    The particle 2 has $d_1=1$ nm and $q_1=+e$. The range of parameters for particle 1
    is the same as in \fref{fig:etacoulsmall} for the Coulomb case.}
  \label{fig:etampcsmall}
  \end{figure}

  \Fref{fig:etampcsmall} shows a chart of the enhancement factor for the same size and charge pairs as those in 
  \fref{fig:etacoulsmall}. It appears clear that most values are increased in comparison
  with \fref{fig:etacoulsmall}. In particular, $\eta_{12}$ for like-charged particles is higher for the
  largest particles, and the neutral-charged interaction for the smallest particles is significantly increased.
\subsection{MCP neutral-charged enhancement factor}\label{sec:neutrals}
\begin{figure}\centering
  \includegraphics[scale=0.45]{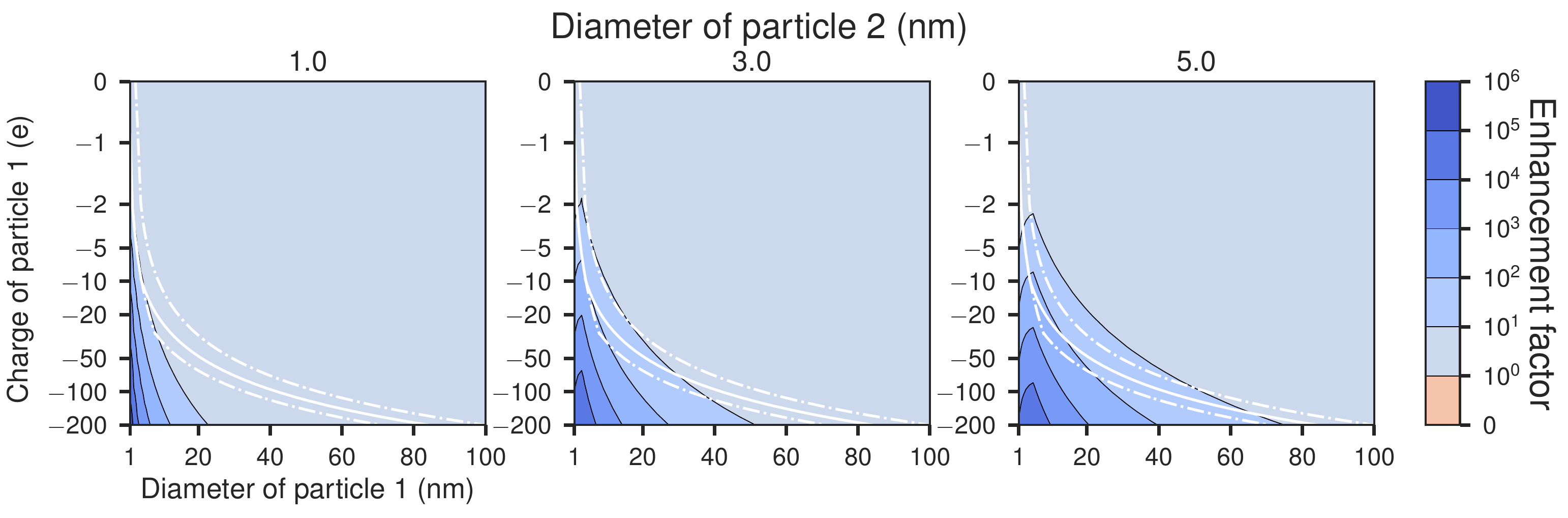}
  \caption{Contour levels of the MCP enhancement factor for a negatively charged particle 1 and a neutral particle 2
  with sizes $d_2=1, 3, and 5$ nm.}
  \label{fig:neutralsmall}
  \end{figure}

\begin{figure}\centering
    \includegraphics[scale=0.45]{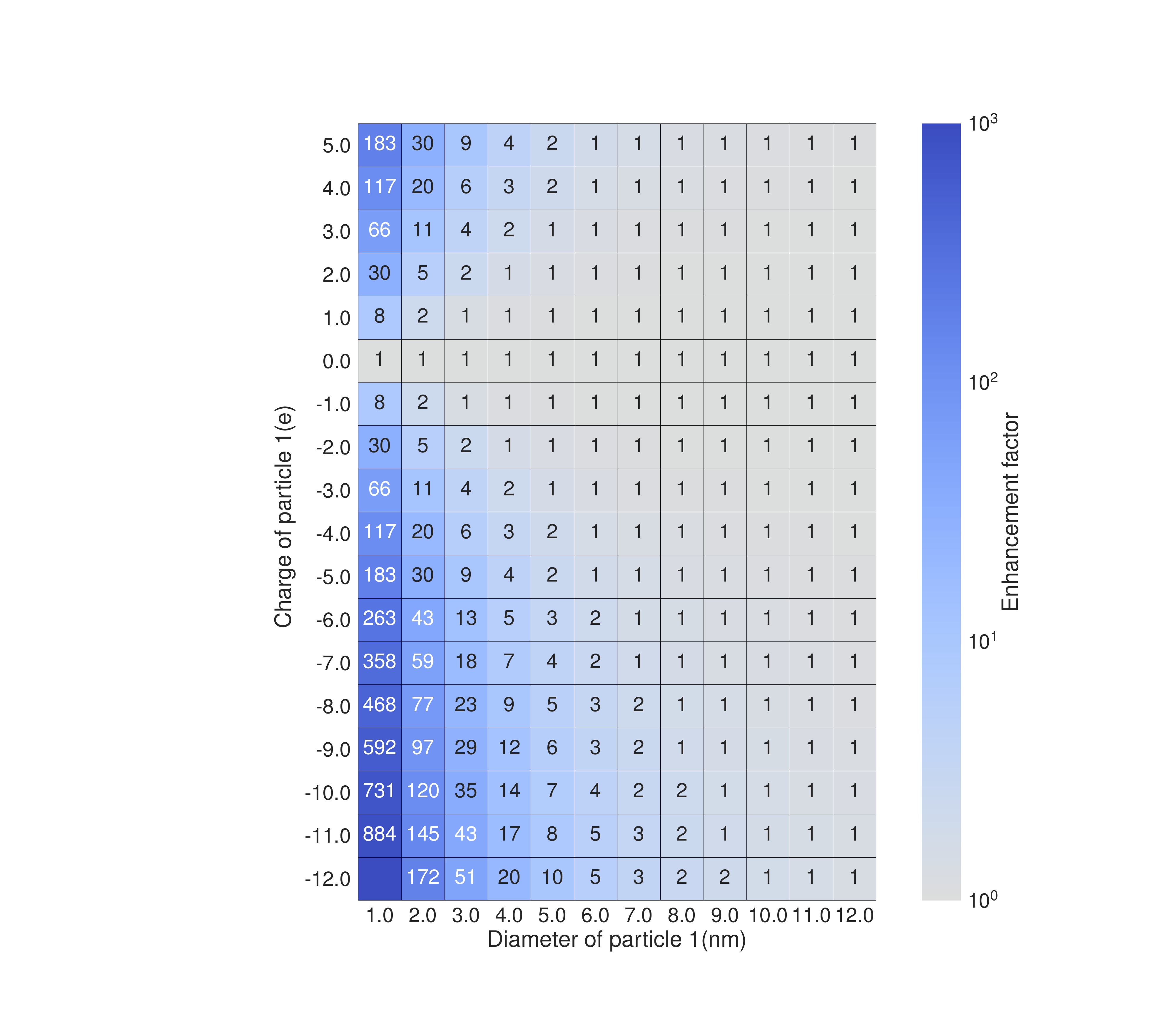}
    \caption{Chart of the MCP enhancement factor for small particles 1 and a neutral particle of
    size $d_2= 1$ nm.}
  \label{fig:etaneutralsmall1}
  \end{figure}

\begin{figure}\centering
    \includegraphics[scale=0.45]{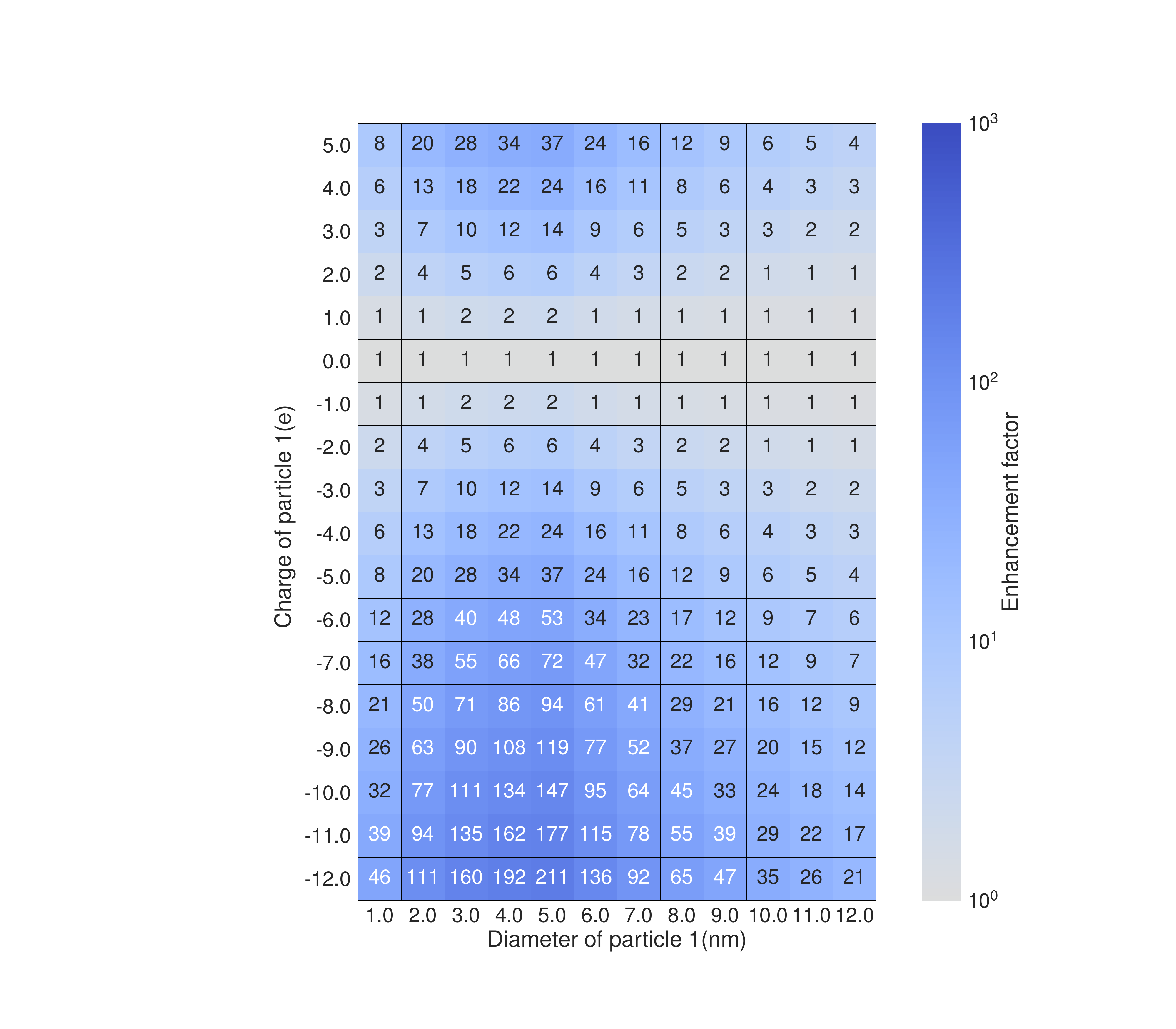}
    \caption{Chart of the MCP enhancement factor for small particles 1 and a neutral particle of
    size $d_2=5$ nm.}
  \label{fig:etaneutralsmall5}
  \end{figure}

  \Fref{fig:neutralsmall} shows the enhancement factor for the interaction between neutral particles 2 of different
  diameters with particles 1 of different charges and sizes.
  It can be seen that $\eta_{12} \ge 1$ for all combinations of ($d_1,q_1$) and $d_2$. 
  Some features of figure \ref{fig:neutralsmall} can be understood by looking at the IPA for $d_2/d_1 \ll 1$,
  \begin{equation}
    \Phi_{\text{IPA},12}(r_{min}) \approx -K \kappa q_1^2
    \frac{d_2^3}{d_1^4} \left( 1-4\frac{d_2}{d_1} \right).
  \end{equation}
  In particular, one can see that $\eta_{12}-1 \propto q_1^2d_2^{3}/d_1^{4}$.
  The charts of figures \ref{fig:etaneutralsmall1} and~\ref{fig:etaneutralsmall5} show the behavior
  of $\eta_{12}$ when $d_1/d_2 \sim 1$.
  One can see that the enhancement factor is larger when the particles have comparable sizes.
  It is interesting to note that this trend is opposite to the behavior of the kernel $\beta_{ij}^0$ shown
  in~\fref{fig:beta_free} where collisions between dissimilar particles are more likely.
 
 Most interestingly, the equilibrium charge of particle 1 lies in the region where enhanced coagulation occurs, as
 can be seen clearly in figure \ref{fig:neutralsmall}.
  Thus, from the coagulation perspective, the most important consequence of induced polarization appears to be the
  increase of the coagulation rate of small particles (which have more chances to be neutral than bigger ones) and
  charged particles.

  By comparing the charts of figures \ref{fig:etampcsmall} and \ref{fig:etaneutralsmall1}, it appears that the
  neutral-charged enhancement is
  always lower than the opposite charge enhancement for the same $d_1$ and $q_1$. The enhancement factors
  for high-charge states of $d_1 =1$ nm can, however, be larger than in the pure Coulomb case
  of~\fref{fig:etacoulsmall}. Such high-charge states are however unlikely for $d_1=1$ nm particles.

\begin{figure}\centering
    \includegraphics[scale=0.45]{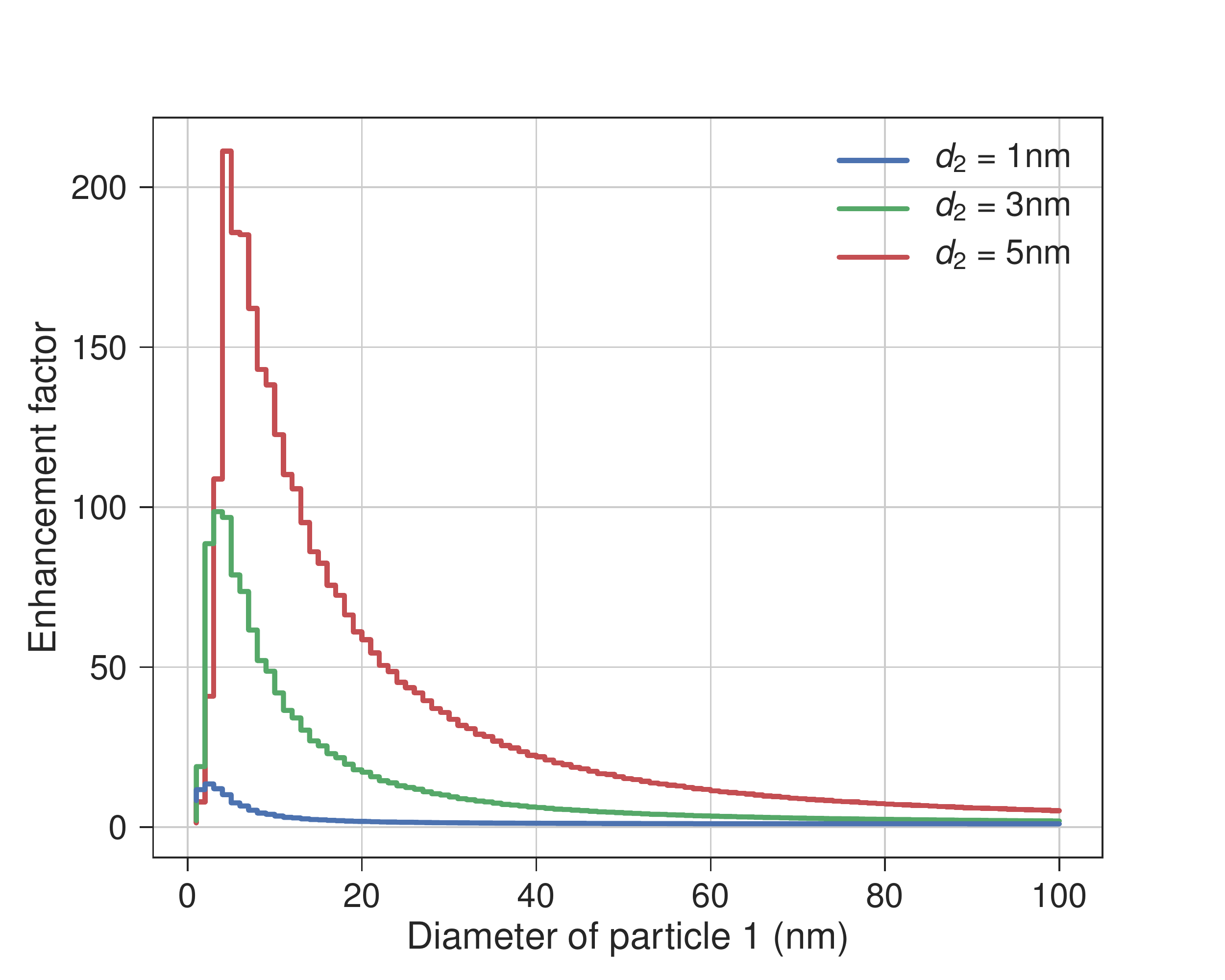}
    \caption{MCP enhancement factor as a function of the size of particle 1, with the 
    corresponding equilibrium charge (\fref{fig:charges}), interacting with neutral particles 2 of 1, 3 and 5 nm.
    The steps arise because of the discreteness of the charge
    and the discretization of the size of particle 1}
  \label{fig:etaneutralcomp}
  \end{figure}

  \Fref{fig:etaneutralcomp} shows the coagulation enhancement factor between particles 1 with their equilibrium
  charge (\fref{fig:charges}) and neutral particles 2 of different sizes, as a function of $d_1$. 
  Since in Coulomb interaction the enhancement factor for a neutral and a charged particle is 1, one sees that induced
  polarization can increase coagulation by one or two orders of magnitude.
  This result suggests that induced polarization could play an essential role in the growth of nanoparticles in dusty
  plasmas.
\section{Conclusion}\label{sec:con}
 This paper revisits the calculation of the coagulation enhancement factor resulting from the electrostatic
 interaction of silicon nanoparticles in
 typical laboratory conditions of low-temperature argon-silane plasmas. The approach used is based on the rigorous
 multipolar coefficient potential (MCP) of Bichoutskaia et al. \cite{bichoutskaia_elena_electrostatic_2010}. In
 contrast to previous investigations using simpler potentials
 forms~\cite{ravi_coagulation_2009, Huang1991191, amadon_cluster-collision_1991, ouyang_nanoparticle_2012},
 the MCP is not singular at the contact. This property allows using the straightforward orbital-motion limited (OML)
 theory to calculate the enhancement factor.
  
  The induced polarization as calculated by the MCP produces an attractive force component when one of the two
  interacting particles is charged. Several significant differences have been observed as compared to the pure Coulomb
  interaction.  In particular, enhancement ($\eta_{ij}>1$), i.e. overall attraction, can be found in like-charged
  interaction. Such cases involve, however, particles far from charge equilibrium in the plasma, as calculated using
  the tunnel electron current and the OML plasma currents.
  In addition, the enhancement in neutral-charged particles interaction provided by the MCP can be
  several orders of magnitude and takes place for particles near charge equilibrium in the plasma. 
  This enhancement is, however, always lower than for the Coulomb attraction for particles close to charge equilibrium,
  but it can be higher for unlikely highly charged particles. 
  Other significant differences with respect to the pure Coulomb interaction have been observed, such as the opposite
  variation of the enhancement factor with the particle sizes for given charges, and the non-monotonous variation of
  the enhancement factor as a function of the charges for given particle sizes.

  Several of the trends observed in our numerical calculation using the MCP have been understood using a simple
  approximate analytical potential, called the image potential approximation (IPA), which provides an useful
  approximation of the MCP at large interparticle distance and when the particle sizes are very different. On the
  computational side, the IPA provides a convenient means to calculate the potential barrier $\Phi_{ij, \text{max}}$
  involved in the enhancement factor, which is a time-consuming operation when using only the MCP.

  The findings of this work could be readily extended to different materials as well as to different laboratory
  conditions, granted that the Debye length is larger than the average distance between nanoparticles and the system
  remains in the free molecular regime. The enhancement factor presented here can be implemented in a dynamic
  coagulation model, involving coupled sectional,
  charging, and plasma chemistry models~\cite{agarwal_sectional_2012}, to describe the nanoparticle charge and size
  distributions as a function of time. Besides, a proper description of the van der Waals interaction 
  would be required. These topics are currently under investigation.
\appendix
\section{Derivation of the coagulation kernel}\label{app:derivrate}
  The coagulation rate for two hard spheres is~\cite{friedlander_smoke_2000}:
  \begin{equation}
    R = \beta_{ij} n_i n_j,
    \label{eq:coagrate2}
  \end{equation}
  where the coagulation kernel $\beta_{ij}$ is the collision frequency defined as the product of the collection
  cross-section and the relative velocity of the two spheres,
  \begin{equation}
    \beta_{ij} = \sigma_{ij}v_{ij},
    \label{eq:betasigmav}
  \end{equation}
  In this expression, $m_{ij} = m_i m_j/(m_i+m_j)$ is the reduced mass for particles $i$ and $j$. Suppose that the
  particle velocity distribution is Maxwellian:
  \begin{equation}
   f_{ij}(v) = 4\pi v^2 \left( \frac{m_{ij}}{2\pi \kboltz T}\right)^{3/2}
              \exp\left( -\frac{m_{ij}v^2}{2 \kboltz T}\right).
    \label{eq:maxwell}
  \end{equation}
  Hence, it is necessary to integrate over all possible velocities,
  \begin{equation}
    \beta_{ij} = \int^\infty_{0} v\sigma_{ij}(v)f_{ij}(v)\text{d}v.
    \label{eq:betaint}
  \end{equation}
  Furthermore, the collection cross section is obtained in terms of the impact parameter~\cite{piel_plasma_physics} as,
  \begin{equation}
   \sigma_{ij} = \pi b^2_{ij},
    \label{eq:crosssection}
  \end{equation}
 In case of a central force, the impact parameter can be derived from laws of conservation of energy
 and angular momentum~\cite{allen_probe_1992}:
  \begin{equation}
   b_{ij} = r_\text{min}\left( 1-\frac{2\Phi_{ij}(r_\text{min})}{m_{ij}v^2} \right)^{1/2}.
    \label{eq:impact}
  \end{equation}
  Then, from \eref{eq:betaint} it follows,
  %
  \begin{equation}
   \beta_{ij} = 4\pi^2 r^2_\text{min} \left( \frac{m_{ij}}{2\pi \kboltz T}\right)^{3/2}
    \int^\infty_{v_\text{min}} \left( 1-\frac{2\Phi_{ij}(r_\text{min})}{m_{ij}v^2} \right)
      \exp\left( -\frac{m_{ij}v^2}{2 \kboltz T}\right)v^3\text{d}v,
    \label{eq:beta1}
  \end{equation}
  %
  where  $v_\text{min}$ is given by,
  \begin{equation}
   v^2_\text{min} = \frac{2\Phi_{ij, \text{max}}}{m_{ij}}, \qquad
    v_\text{min} = \Re\left[ \left( \frac{2\Phi_{ij, \text{max}}}{m_{ij}}\right)^{1/2} \right].
    \label{eq:vmin}
  \end{equation}
 is the minimum relative velocity required to breach the
  potential barrier (if such a barrier exists). Rearranging the integral~\eref{eq:beta1} gives,
  %
  \begin{equation}
   \beta_{ij} = \left( \frac{8\pi \kboltz T}{m_{ij}}\right)^{1/2}\left(r_i + r_j\right)^2
    2\int^\infty_{\tilde{v}_\text{min}} \left( 1-\frac{\Phi_{ij}(r_\text{min})}{\kboltz T v^2} \right)
      \exp\left( -v^2\right)v^3\text{d}v,
    \label{eq:beta2}
  \end{equation}
  %
  with $\tilde{v}_\text{min}=\Re\left[ \left( \frac{\Phi_{ij, \text{max}}}{\kboltz T}\right)^{1/2} \right]$.
  Now it is possible to identify the coagulation kernel for hard spheres as~\cite{friedlander_smoke_2000},
  \begin{equation}
   \beta^0_{ij} = \left( \frac{8\pi \kboltz T}{m_{ij}}\right)^{1/2}\left(r_i + r_j\right)^2,
    \label{eq:beta0}
  \end{equation}
  This result is equivalent to~\eref{eq:neutralkernel} but expressed
  in terms of the radii of the colliding particles. Thus, the enhancement factor
  is~\cite{ouyang_nanoparticle_2012}:
  \begin{equation}
   \eta_{ij} = 2\int^\infty_{\tilde{v}_\text{min}} \left( 1-\frac{\Phi_{ij}(r_\text{min})}{\kboltz T v^2} \right)
      \exp\left( -v^2\right)v^3\text{d}v,
    \label{eq:etaint}
  \end{equation}
  %
  Finally, after integration, one obtains:
  \begin{equation}
   \eta_{ij} = \exp\left(-\frac{\Phi_{ij, \text{max}}}{\kboltz T}\right)
      \left[ 1 + \frac{\Phi_{ij, \text{max}}-\Phi_{ij}(r_\text{min})}{\kboltz T} \right].
    \label{eq:etagen2}
  \end{equation}
\section{Image Potential Approximation (IPA)}\label{app:mpc-ipa-comparison}
  As mentioned in~\sref{sec:num}, using the MCP to compute the enhancement factor is generally time-consuming.
  Moreover, it is difficult to draw general conclusions from this formulation. Thus, it would be desirable
  to find an alternative
  description for the electrostatic interaction which retains most features of the MCP.
  Draine \& Sutin~\cite{draine_collisional_1987} derived the following simple expression for the image potential of
  a point charge $q_j$ and a sphere of radius $r_i$ and charge $q_i$,
  \begin{align}
    \Phi_{\text{DS}}(r, r_i, q_i,q_j) = K\frac{q_i q_j}{r}
  -\frac{K\kappa q^2_j r^3_i}{2r^2\left( r^2-r^2_i\right)}.
  \label{eq:IP}
  \end{align}
  where $\kappa = (\varepsilon -1)/(\varepsilon + 2)$. This potential corresponds to the classical Coulomb interaction 
  potential energy plus a correction corresponding to the image
  potential due to the dielectric nature of the spherical particle.

  For the problem of interest here, it is not possible to assume a point charge particle. 
  Therefore, a more suitable expression would include the image potential energy of the
  nanoparticle $i$ in the nanoparticle $j$:
  \begin{align}
  \Phi_\text{IPA}(r, r_i, r_j,q_i,q_j) \equiv \Phi_\text{IPA,ij}
    = K\frac{q_i q_j}{r}
    -\frac{K \kappa q^2_i r^3_j}{2r^2\left( r^2-r^2_j\right)}
    -\frac{K\kappa q^2_j r^3_i}{2r^2\left( r^2-r^2_i\right)}.
  \label{eq:IPA}
  \end{align}
 We call this formulation the Image Potential Approximation (IPA). From equation (\ref{eq:IPA}),
  it is easy to find an analytical expression for the force $F_\text{IPA,ij}=-d\Phi_\text{IPA,ij}/dr$ which
  can be used to find $\Phi_{ij, \text{max}}$, as required to calculate the enhancement factor.

  One can see readily that, in the case of spheres of same charge polarity, the IPA~\eref{eq:IPA} is composed of
  a repulsive Coulomb term plus two attractive contributions. As a result, a competition
  between the long-range Coulomb repulsion and the short-range image potential attraction creates
  the potential barrier $\Phi_{ij, \text{max}}$, which can be breached only if the nanoparticles have sufficient
  kinetic energy.
  For the neutral-charged interaction there is no Coulomb contribution; hence 
  enhanced coagulation always takes place.

\begin{figure}\centering
  \includegraphics[scale=0.6]{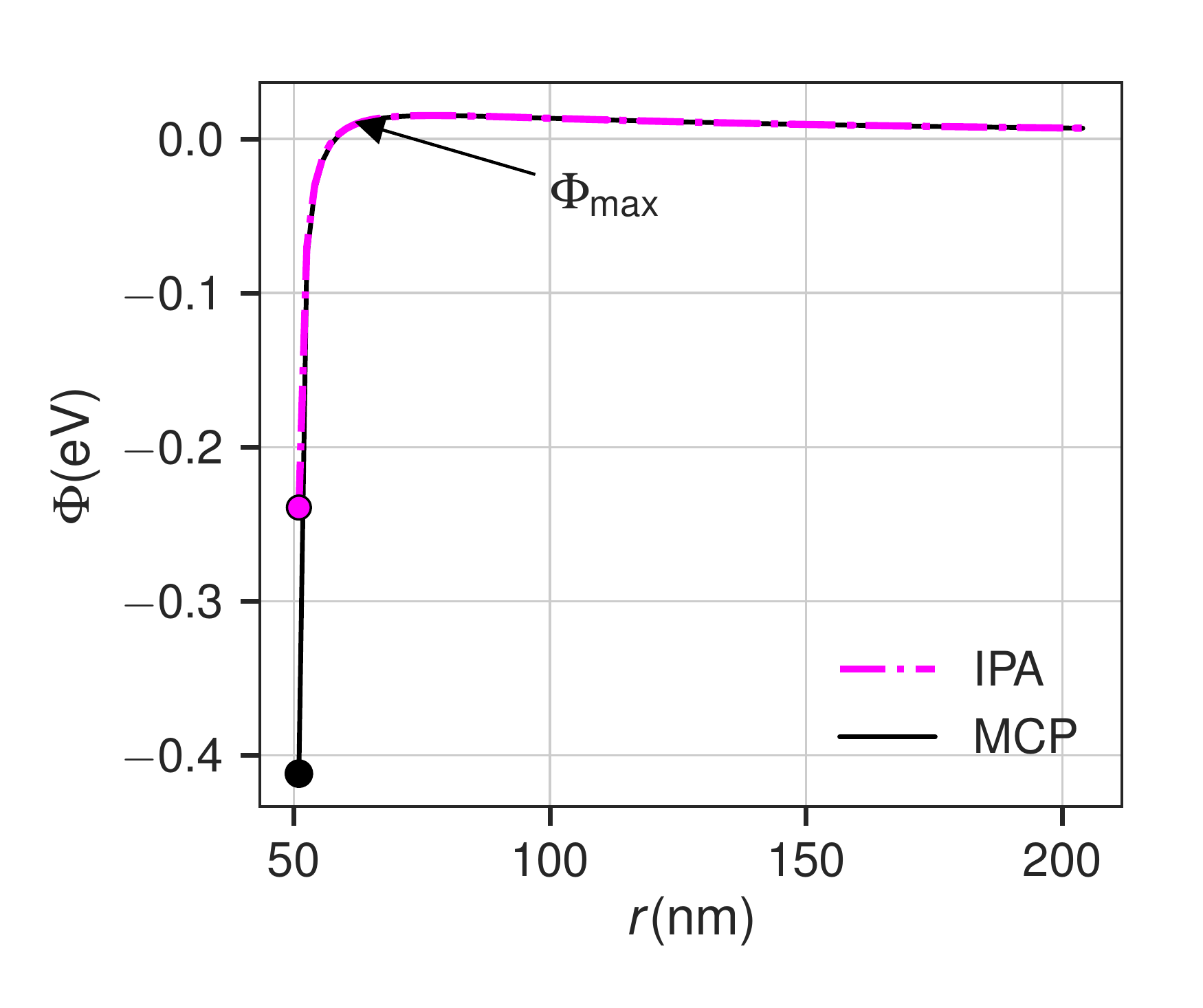}
  \caption{Potential energies as a function of the interparticle distance when
  $r_1 = \SI{1}{\nano\metre},\, r_2 = \SI{50}{\nano\metre}$, and $q_1 = q_2 = -e$, where $\Phi_{\text{max}}$ is
  the height of the potential barrier.}
  \label{fig:pot_comp}
  \end{figure}

\begin{figure}\centering
    \includegraphics[scale=0.45]{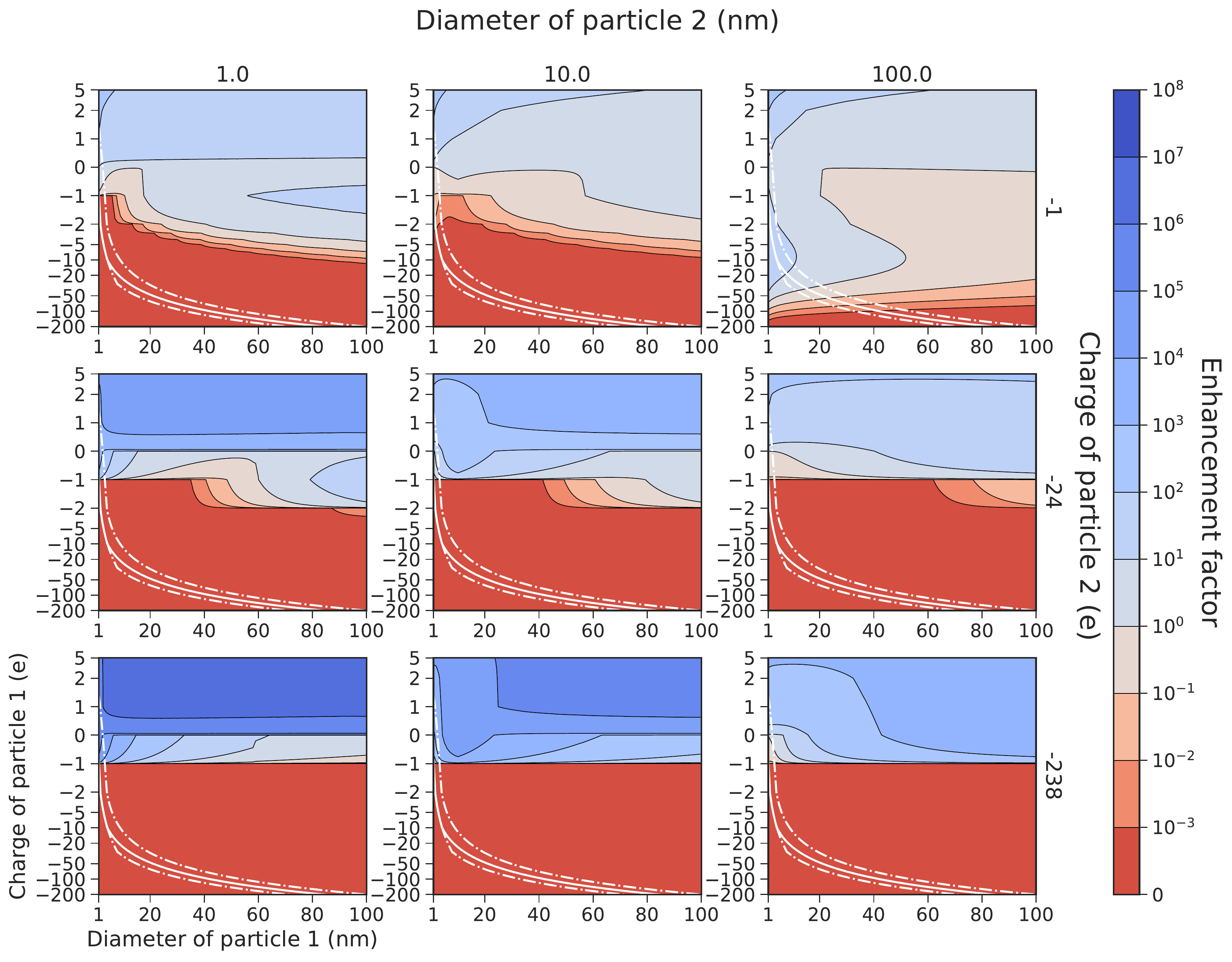}
    \caption{Contour levels of the enhancement factor for the hybrid approach for the same values as 
    in figures \ref{fig:etacoul} and \ref{fig:etapmc}. These
    results are in qualitative agreement with~\fref{fig:etapmc}.}
  \label{fig:etahybrid}
  \end{figure}

  \Fref{fig:pot_comp} shows a comparison between the IPA and the MCP in a case where there is a potential barrier whose
  maximum occurs at $r=r_\text{max}$. 
  The IPA nicely reproduces the long-range behavior of the MCP. This is computationally advantageous because 
  the potential barrier can be located using the simple IPA, and then the MCP  can be
  computed at that location, $r_{\text{max}}$. This procedure proves to be 15 to 20 times faster than finding
  $r_{\text{max}}$ directly from the MCP. Figure \ref{fig:etahybrid} shows the enhancement factor
  using this hybrid approach: the location of the barrier was determined using the IPA force, and then
  the height of the barrier and the potential at contact were computed using the MCP formulation. One can
  hardly see any difference between figures \ref{fig:etahybrid} and \ref{fig:etapmc}.
\section{Image Potential Approximation (IPA) Barrier}\label{app:maximum-image-potential-approximation-ipa}
  The purposes of this Appendix are to (i) write out the equation used to calculate $r_{\text{max}}$, i.e., the
  location of the maximum of the potential, from the IPA prescription,
  (ii) show graphical solutions in particular cases, and (iii) show that $r_{\text{max}}$
  is on the order of magnitude of the particles size, and thus much smaller than the Debye length
  in the conditions stated in~\tref{tab:experiments}.

\begin{figure}\centering
  \includegraphics[scale=0.6]{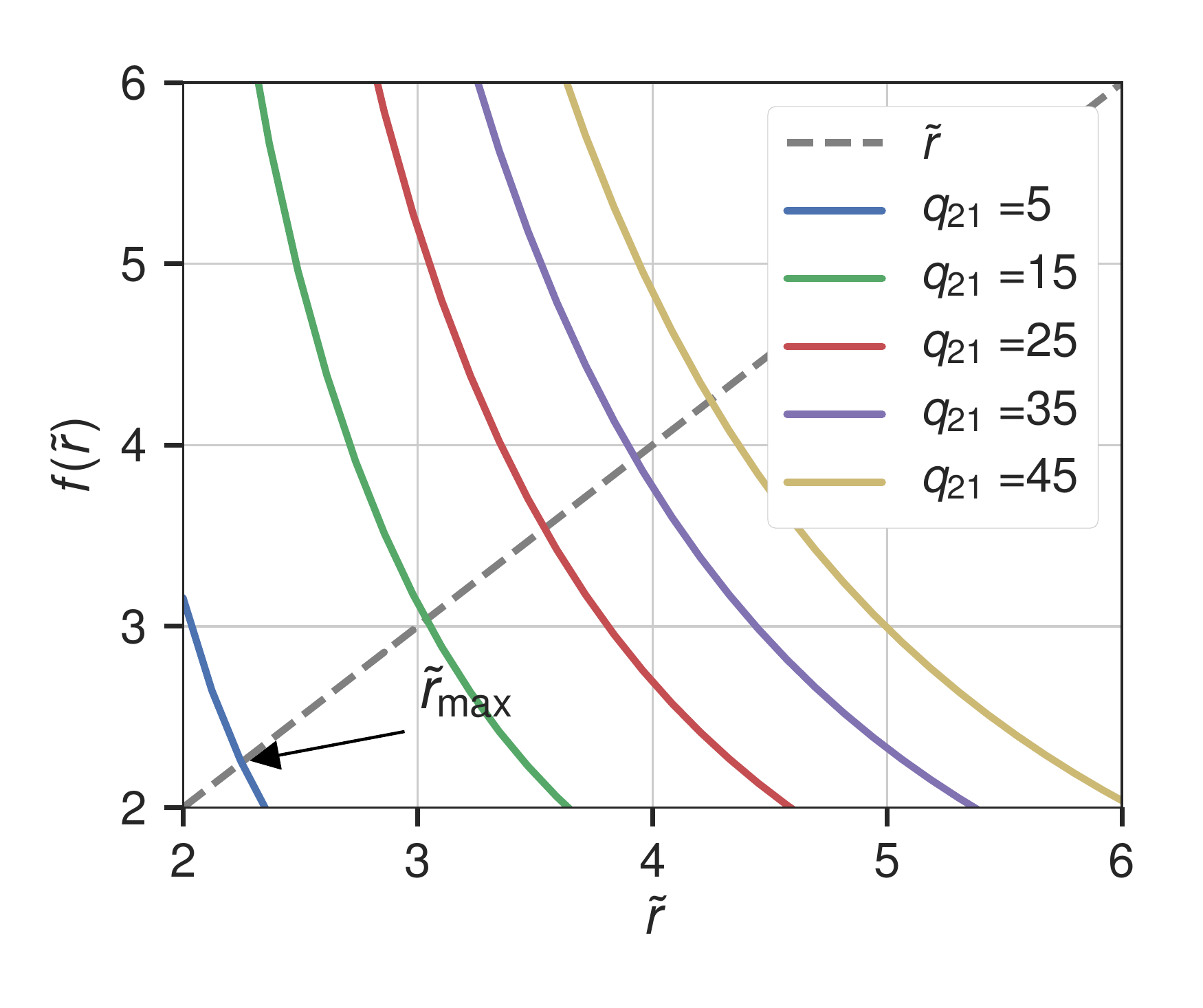}
  \caption{Graphical solution of equation~\eref{eq:rmax1}. The intersection of the dashed line and the 
  full lines gives $\tilde{r}_\text{max}$
  which is near the contact radius $\tilde{r}=2$.}
  \label{fig:rmax1}
  \end{figure}

  Equation (\ref{eq:IPA}) can be expressed in the dimensionless form,
  \begin{align}
    \hat{\Phi}_\text{IPA,12} = 
    \frac{q_{21}}{\tilde{r}}
    -\frac{\kappa r^3_{21}}{2\tilde{r}^2\left( \tilde{r}^2-r^2_{21}\right)}
    -\frac{\kappa q^2_{21}}{2\tilde{r}^2\left( \tilde{r}^2-1\right)},
  \end{align}
  where 
\(\tilde{r}=r/r_1\), \(r_{21}=r_2/r_1\), and \(q_{21}=q_2/q_1\).
  For particles of same polarity \(q_{21}>0\) the associated dimensionless force is,
  \begin{equation}
    \hat{F}_\text{IPA,12} 
    = -\Deriv{\hat{\Phi}_\text{IPA,12}}{\tilde{r}},
    \label{eq:fipa}
  \end{equation}
  which is equal to zero at the potential barrier located at \(\tilde{r}=\tilde{r}_\text{max}\),
  \begin{align}
    \tilde{r}_\text{max} = 
    \left.\kappa\frac{r^3_{21}\left( 2\tilde{r}^2-r^2_{21}\right)}{q_\text{21}\left( \tilde{r}^2-r^2_{21}
    \right)^2}
    \right|_{\tilde{r}=\tilde{r}_\text{max}}
   +\left.\kappa\frac{q_{21}\left( 2\tilde{r}^2-1\right)}{\left( \tilde{r}^2-1\right)^2}
    \right|_{\tilde{r}=\tilde{r}_\text{max}}.
    \label{eq:rmax}
  \end{align}
  A simple case arises for particles of the same size, i.e., $r_\text{21}=1$:
  \begin{align}
    \tilde{r}_\text{max} = \kappa
    \frac{(1+q^2_{21})}{q_{21}}
    \frac{\left( 2\tilde{r}^2_\text{max}-1\right)}{\left( \tilde{r}^2_\text{max}-1\right)^2}.
    \label{eq:rmax1}
  \end{align}
  As shown in~\fref{fig:rmax1}, the solution of equation (\ref{eq:rmax1}) is the value of $\tilde{r}$ 
  at the intersection between the
  dotted line (left-hand side of (\ref{eq:rmax1})) and the full lines (right-hand side of (\ref{eq:rmax1}))
  corresponding to various values of $q_{21}$.
  It can be seen that a solution is found for each value of $q_{21}$ beyond the contact radius 
  $\tilde{r}>1+r_{21}=2$. Figure \ref{fig:rmax1} also shows that  $\tilde{r}_\text{max}$ is on the order of 
  magnitude of the size of the particles and thus close to the contact radius.

\begin{figure}\centering
  \includegraphics[scale=0.6]{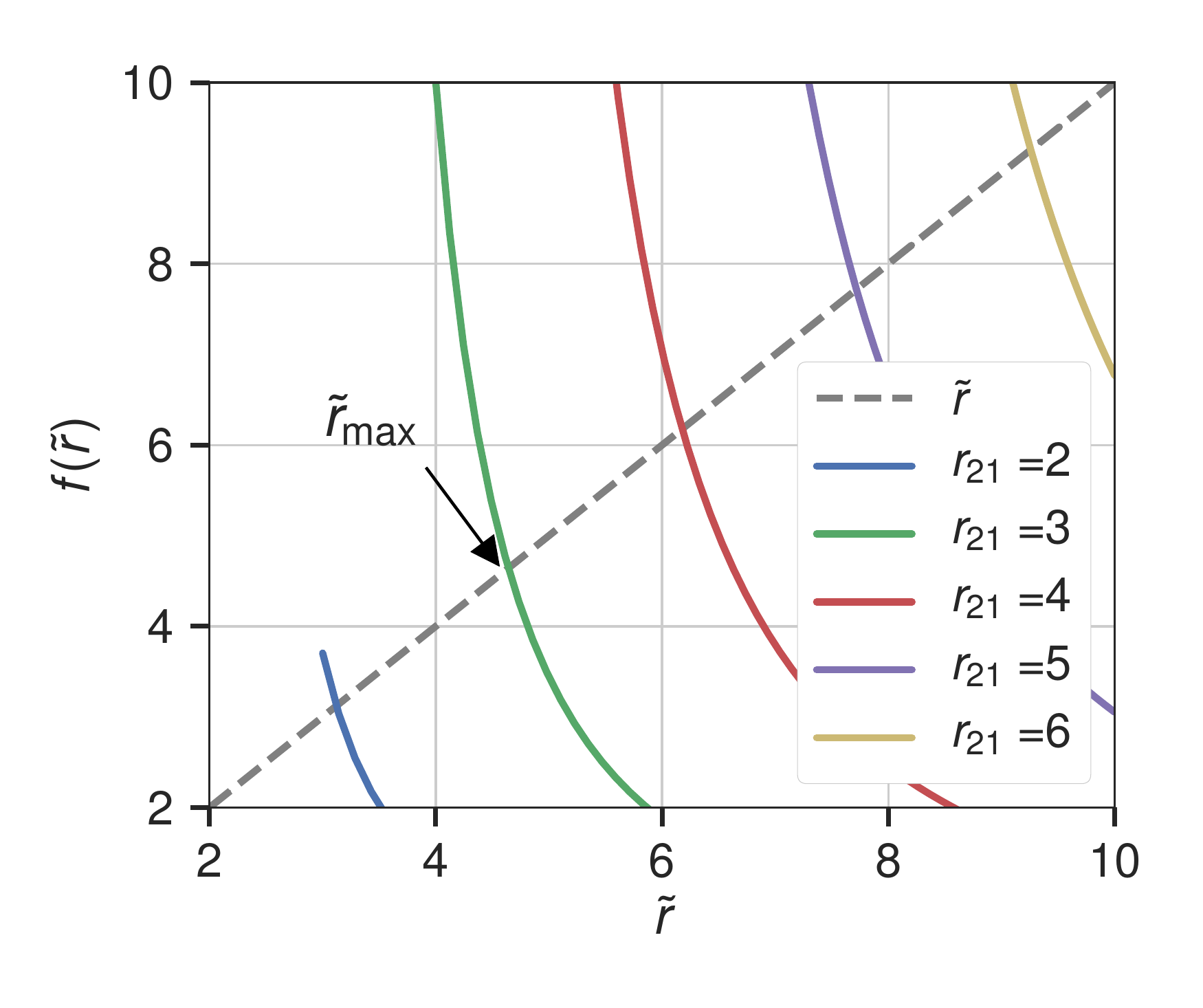}
  \caption{Graphical solution of equation~\eref{eq:rmax} for equally charged nanoparticles. For the selected
  charge ratios $q_{21}$, the solutions $\tilde{r}_\text{max}$ is near the contact radius $\tilde{r}=1+r_{21}$.}
  \label{fig:rmax2}
  \end{figure}

\begin{figure}\centering
  \includegraphics[scale=0.6]{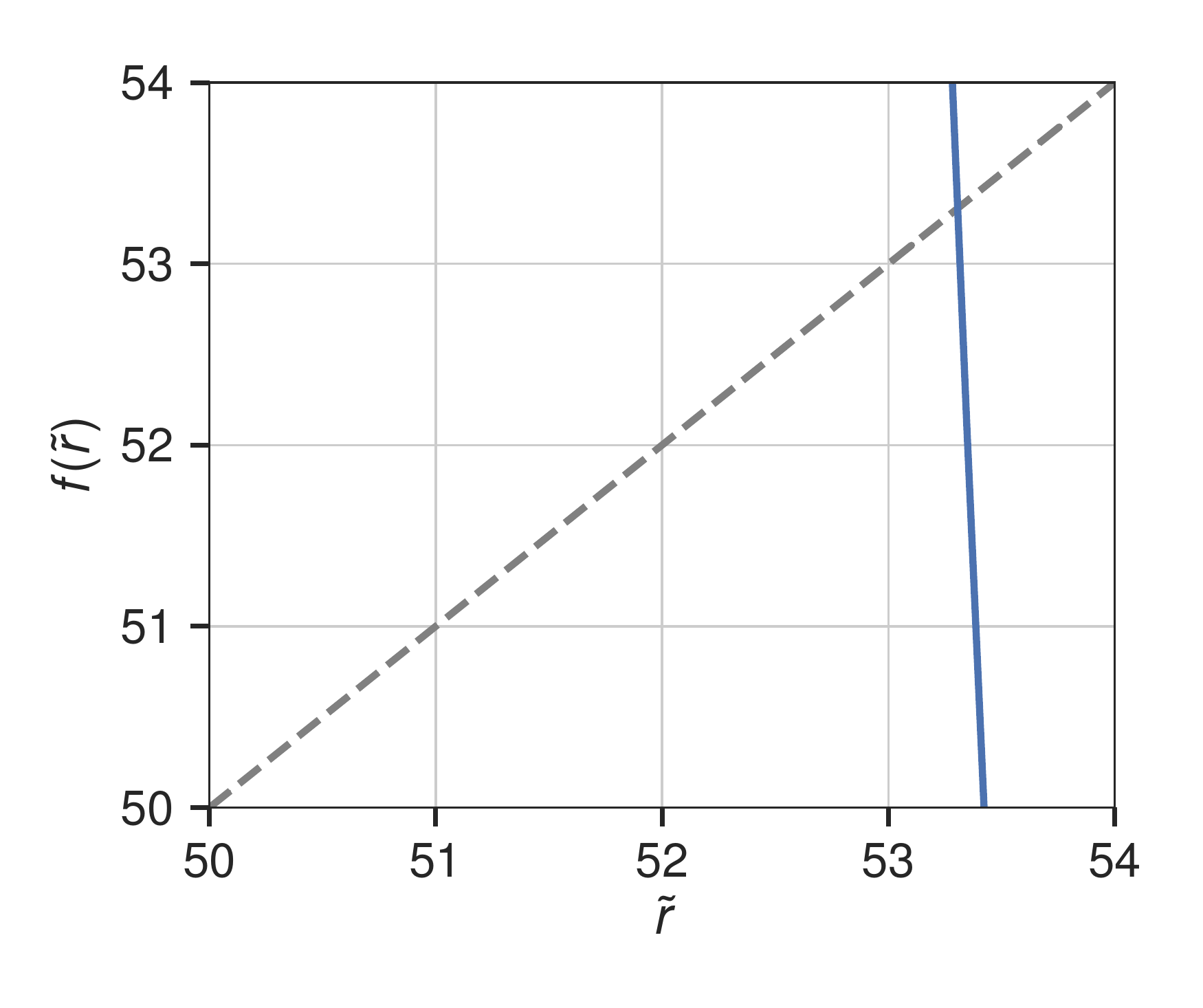}
  \caption{Graphical solution of equation~\eref{eq:rmax} for high charge and size ratios
  $r_{21}=q_{21}=50$.}
  \label{fig:rmax3}
  \end{figure}

  The analysis can be repeated for the other simple cases of equally charged spheres $q_{21}=1$ and
  different size ratios $r_{21}$. The graphical solution of equation (\ref{eq:rmax}) in such cases is shown
  in~\fref{fig:rmax2}.
  Same as in the previous case where $r_{21}=1$, the solutions $r_\text{max}$ are on the same order of 
  magnitude as the particles size.
 The same conclusion can be drawn in the case of high charge and size ratios $r_{21}=q_{21}=50$ as shown 
 in~\fref{fig:rmax3}. 
 
 The fact that $r_\text{max}$ is similar to the particles size is due to the fast decrease
 in $r^{-5}$ as $(r/r_k)^2 \gg 1$ of the induced polarization force, which must produce a zero of the force at short
 separation distance, if such a zero is to exist.
 
\newpage
\section*{References}
\begin{thebibliography}{10}

\bibitem{boufendi_particle_1994}
L.~Boufendi and A.~Bouchoule.
\newblock Particle nucleation and growth in a low–pressure argon–silane
  discharge.
\newblock {\em Plasma Sources Sci. Technol.}, 3(3):262, August 1994.

\bibitem{hollenstein_physics_2000}
Ch~Hollenstein.
\newblock The physics and chemistry of dusty plasmas.
\newblock {\em Plasma Phys. Control. Fusion}, 42(10):R93, October 2000.

\bibitem{boufendi_dusty_2011}
L.~Boufendi, M.~Ch Jouanny, E.~Kovacevic, J.~Berndt, and M.~Mikikian.
\newblock Dusty plasma for nanotechnology.
\newblock {\em J. Phys. D: Appl. Phys.}, 44(17):174035, May 2011.

\bibitem{fortov_complex_2005}
V.~E. Fortov, A.~V. Ivlev, S.~A. Khrapak, A.~G. Khrapak, and G.~E. Morfill.
\newblock Complex (dusty) plasmas: Current status, open issues, perspectives.
\newblock {\em Physics Reports}, 421(1–2):1--103, December 2005.

\bibitem{mamunuru_existence_2017}
M.~Mamunuru, R.~Le Picard, Y.~Sakiyama, and S.~L. Girshick.
\newblock The {Existence} of {Non}-negatively {Charged} {Dust} {Particles} in
  {Nonthermal} {Plasmas}.
\newblock {\em Plasma Chem Plasma Process}, 37(3):701--715, May 2017.

\bibitem{lindgren_progress_2016}
Eric~B. Lindgren, Ho-Kei Chan, Anthony~J. Stace, and Elena Besley.
\newblock Progress in the theory of electrostatic interactions between charged
  particles.
\newblock {\em Physical Chemistry Chemical Physics}, 18(8):5883--5895, February
  2016.

\bibitem{ravi_coagulation_2009}
L.~Ravi and S.~L. Girshick.
\newblock Coagulation of nanoparticles in a plasma.
\newblock {\em Phys. Rev. E}, 79(2):026408, February 2009.

\bibitem{Huang1991191}
David~D Huang, John~H Seinfeld, and Kikuo Okuyama.
\newblock Image potential between a charged particle and an uncharged particle
  in aerosol coagulation-enhancement in all size regimes and interplay with van
  der waals forces.
\newblock {\em Journal of Colloid and Interface Science}, 141(1):191 -- 198,
  1991.

\bibitem{amadon_cluster-collision_1991}
A.~S. Amadon and W.~H. Marlow.
\newblock Cluster-collision frequency. {II}. {Estimation} of the collision
  rate.
\newblock {\em Physical Review A}, 43(10):5493--5499, May 1991.

\bibitem{ouyang_nanoparticle_2012}
Hui Ouyang, Ranganathan Gopalakrishnan, and Christopher J.~Hogan Jr.
\newblock Nanoparticle collisions in the gas phase in the presence of singular
  contact potentials.
\newblock {\em The Journal of Chemical Physics}, 137(6):064316, August 2012.

\bibitem{bichoutskaia_elena_electrostatic_2010}
{Bichoutskaia, Elena}, {Boatwright, Adran L.}, {Khachatourian, Armik}, and
  {Stace, Anthony J.}
\newblock Electrostatic analysis of the interactions between charged particles
  of dielectric materials.
\newblock {\em The Journal of Chemical Physics}, 133(2):024105, July 2010.

\bibitem{allen_probe_1992}
J.~E. Allen.
\newblock Probe theory - the orbital motion approach.
\newblock {\em Phys. Scr.}, 45(5):497, May 1992.

\bibitem{Hamaker_1937}
H.~C. Hamaker.
\newblock The {L}ondon-van der {W}aals attractino between spherical particles.
\newblock {\em Physica IV}, 10:1058, November 1937.

\bibitem{Priye_2013}
Aashish Priye and William~H Marlow.
\newblock Computations of {L}ifshitz–van der {W}aals interaction energies
  between irregular particles and surfaces at all separations for resuspension
  modelling.
\newblock {\em J. Phys. D: Appl. Phys.}, 46:425306, 2013.

\bibitem{shukla_colloquium_2009}
P.~Shukla.
\newblock Colloquium: Fundamentals of dust-plasma interactions.
\newblock {\em Rev. Mod. Phys.}, 81(1):25--44, 2009.

\bibitem{fortov_dusty_2004}
V.~E. Fortov, A.~G. Khrapak, S.~A. Khrapak, V.~I. Molotkov, and Petrov Petrov.
\newblock Dusty plasmas.
\newblock {\em Physics - Uspekhi}, 47:447--492, 2004.

\bibitem{picard_effect_2016}
Romain~Le Picard and Steven~L. Girshick.
\newblock The effect of single-particle charge limits on charge distributions
  in dusty plasmas.
\newblock {\em Journal of Physics D: Applied Physics}, 49(9):095201, 2016.

\bibitem{haynes2016crc}
W.M. Haynes.
\newblock {\em CRC Handbook of Chemistry and Physics, 97th Edition}.
\newblock CRC Press, 2016.

\bibitem{friedlander_smoke_2000}
Sheldon~Kay Friedlander.
\newblock {\em Smoke, Dust, and Haze: Fundamentals of Aerosol Dynamics}.
\newblock Oxford University Press, 2000.

\bibitem{heijmans_comment_2016}
L.~C.~J. Heijmans, F.~M. J. H. van~de Wetering, and S.~Nijdam.
\newblock Comment on ‘{The} effect of single-particle charge limits on charge
  distributions in dusty plasmas’.
\newblock {\em Journal of Physics D: Applied Physics}, 49(38):388001, 2016.

\bibitem{griffiths_wkb}
D.~J. Griffiths.
\newblock {\em {Introduction to Quantum Mechanics}}, chapter {The WKB
  Approximation}, pages 315--339.
\newblock Pearson Prentice Hall, Upper Saddle River, NJ, 2nd edition, 2005.

\bibitem{stace_reply_2012}
{Stace, Anthony J.} and {Bichoutskaia, Elena}.
\newblock Reply to the ‘{Comment} on “{Treating} highly charged carbon and
  fullerene clusters as dielectric particles”’ by {H}. {Zettergren} and
  {H}. {Cederquist}, {Phys}. {Chem}. {Chem}. {Phys}., 2012, 14, {DOI}:
  10.1039/c2cp42883k.
\newblock {\em Physical Chemistry Chemical Physics}, 14(48):16771--16772,
  November 2012.

\bibitem{khachatourian_electrostatic_2014}
Armik Khachatourian, Ho-Kei Chan, Anthony~J. Stace, and Elena Bichoutskaia.
\newblock Electrostatic force between a charged sphere and a planar surface:
  {A} general solution for dielectric materials.
\newblock {\em The Journal of Chemical Physics}, 140(7):074107, February 2014.

\bibitem{boostmath}
John Maddock, Paul Bristow, Hubert Holin, Xiaogang Zhang, Bruno Lalande, Johan
  Rade, Gautam Sewani, and Thijs Van~Den Berg.
\newblock {\em Boost C++ Math Toolkit}.
\newblock hetp, 2009.

\bibitem{scipy}
Eric Jones, Travis Oliphant, Pearu Peterson, et~al.
\newblock {SciPy}: Open source scientific tools for {Python}, 2001--.

\bibitem{Hunter_2007}
J.~D. Hunter.
\newblock Matplotlib: A 2d graphics environment.
\newblock {\em Computing In Science \& Engineering}, 9(3):90--95, 2007.

\bibitem{agarwal_sectional_2012}
P.~Agarwal and S.L. Girshick.
\newblock Sectional modeling of nanoparticle size and charge distributions in
  dusty plasmas.
\newblock {\em Plasma Sources Sci. Technol.}, 21(5):055023, October 2012.

\bibitem{piel_plasma_physics}
Alexander Piel.
\newblock {\em Plasma Physics - An Introduction to Laboratory, Space, and
  Fusion Plasmas}.
\newblock Springer Berlin Heidelberg, January 2010.

\bibitem{draine_collisional_1987}
B.~T. Draine and Brian Sutin.
\newblock Collisional charging of interstellar grains.
\newblock {\em The Astrophysical Journal}, 320:803, September 1987.

\end{thebibliography}
\end{document}